\documentclass[11pt]{article}

\usepackage[left=1.2in,right=1.2in,top=1.2in,bottom=1.2in]{geometry}
\usepackage{hyperref}
\usepackage{graphicx}
\usepackage{amssymb}
\usepackage{amsbsy}
\usepackage{amsmath}
\usepackage{hyperref}
\usepackage{xcolor}
\hypersetup{
    colorlinks,
    linkcolor={red!50!black},
    citecolor={blue!50!black},
    urlcolor={blue!80!black}
}

\newcommand{\rf}[1]{(\ref{#1})}
\newcommand{\beq}{\begin{equation}}
\newcommand{\eeq}{\end{equation}}
\newcommand{\be}{\begin{equation}}
\newcommand{\ee}{\end{equation}}
\newcommand{\bea}{\begin{eqnarray}}
\newcommand{\eea}{\end{eqnarray}}
\newcommand{\eq}[1]{eq.~(\ref{#1})}
\newcommand{\non}{\nonumber \\*}
\newcommand{\ie}{{i.e.}\ }
\newcommand{\bv}{\begin{verbatim}}
\newcommand{\ev}{\end{verbatim}}

\def\Xint#1{\mathchoice
   {\XXint\displaystyle\textstyle{#1}}%
   {\XXint\textstyle\scriptstyle{#1}}%
   {\XXint\scriptstyle\scriptscriptstyle{#1}}%
   {\XXint\scriptscriptstyle\scriptscriptstyle{#1}}%
   \!\int}
\def\XXint#1#2#3{{\setbox0=\hbox{$#1{#2#3}{\int}$}
     \vcenter{\hbox{$#2#3$}}\kern-.5\wd0}}

\def\dashint{\Xint-}

\newcommand{\e}{\,\mbox{e}}
\renewcommand{\d}{{\rm d}}
\renewcommand{\i}{{\rm i}}

\newcommand{\sqL}{\sqrt{\Lambda} }

\newcommand{\tr}{\mathrm{tr}}
\newcommand{\LA}{\left\langle}
\newcommand{\RA}{\right\rangle}

\newcommand{\arctanh}{\,{\rm arctanh}\,}

\newcommand{\tV}{\tilde{V}}
\newcommand{\tU}{\tilde{U}}
\newcommand{\hV}{\hat{V}}

\def\fun#1#2{\lower3.6pt\vbox{\baselineskip0pt\lineskip.9pt
\ialign{$\mathsurround=0pt#1\hfil##\hfil$\crcr#2\crcr\sim\crcr}}}

\begin{document}

\hfill  ITEP--TH--33/15
\begin{center}
\vspace{24pt}
{ \large \bf 
Generalized multicritical one-matrix models
}

\vspace{30pt}

{\sl J. Ambj\o rn}$\,^{a,b}$,  {\sl T. Budd}$\,^a$
and {\sl Y. Makeenko}$\,^{a,c}$

\vspace{24pt}
{\footnotesize

$^a$~The Niels Bohr Institute, Copenhagen University\\
Blegdamsvej 17, DK-2100 Copenhagen, Denmark.\\
{ email: ambjorn@nbi.dk, budd@nbi.dk}\\

\vspace{10pt}

$^b$~IMAPP, Radboud University, \\
Heyendaalseweg 135, 6525 AJ, Nijmegen, The Netherlands\\

\vspace{10pt}

$^c$~Institute of  Theoretical and Experimental Physics\\
B. Cheremushkinskaya 25, 117218 Moscow, Russia.\\
{ email: makeenko@nbi.dk}\\

}
\end{center}
\vspace{48pt}

\begin{center}
{\bf Abstract}
\end{center}
We show that there exists a simple generalization of Kazakov's  multicritical
one-matrix model, which interpolates between the various multicritical 
points of the model. The associated multicritical potential takes the form of a power series with a heavy tail, leading to a cut of the potential and its derivative at the real axis,
and reduces to a polynomial at Kazakov's multicritical points.
From the combinatorial point of view the generalized model allows polygons of arbitrary large degrees (or vertices
of arbitrary large degree, when considering the dual graphs), and it is the weight 
assigned to these large order polygons which brings about the interpolation between the multicritical points in the one-matrix model.


\newpage

\section{Introduction}

Matrix models have been among the most important tools when discussing 
non-critical strings or 2d quantum gravity coupled to conformal field
theories with central charge $c < 1$. The main interest in the gravitational aspect
came from attempts to non-perturbatively regularize the Polyakov path integral
in spacetime dimension different from 26 \cite{adf,david1,kkm,kkm1}. While the stringy aspect of this program partly failed for physical target space dimensions, the 2d gravity aspect was a very fruitful area of research, initiated in \cite{david2,kkm},
and getting full attention after the seminal paper \cite{kazakov2} by Kazakov. The
latter used the Hermitian matrix model in the large $N$ limit to describe certain 
matter fields interacting with 2d quantum gravity. Eventually  it 
was understood that the models in \cite{kazakov2} describe 2d quantum gravity 
coupled to $(2,2m-1)$ conformal field theories, $m=2,3,\ldots$ \cite{staudacher} (see e.g. \cite{review} for a review). The susceptibility
exponents of these theories were calculated (in a way we will discuss below) to be given by
\beq\label{ja1}
\gamma_s = -\frac{1}{m}.
\eeq
To obtain exponents corresponding to other conformal field theories one 
had to consider multi-matrix models \cite{kazkov,Kos89,KS}. In this paper we will show 
that one can in fact obtain the full range of exponents $\gamma_s\in\mathopen{]}-\infty,0\mathclose{[}$, in the 
large-$N$ limit of the standard one-cut Hermitian matrix model by allowing for potentials with ``heavy tails''.
In the range $s\in\mathopen{]}3/2,5/2\mathclose{[}$ these matrix models have a combinatorial interpretation in terms of random plane graphs (or random planar maps) with high degree vertices or polygon, which have been of recent interest in the mathematical (physics) literature \cite{MLeG11,BBG12,BC16}. 

The rest of this paper is organized as follows. In Sec. \ref{multicritical} we remind
the reader of the multicritical matrix model introduced in \cite{kazakov2}. In Sec. \ref{general} we generalize the results of Sec. \ref{multicritical}, such that any critical
exponent $\gamma_s < 0$ can occur.  The  corresponding potential $V_s(x)$ as 
well as its derivative $V'_s(x)$ are infinite power series in $x$ with cuts on the real axis.
We suggest how one can associate a central charge $c(s)$ to each $s$.
In Sec.\  \ref{riemann-hilbert} we show that the standard way of solving the saddle point equation is still valid. Next we address the question 
of universality (Sec. \ref{universality}) and the corresponding continuum limit 
(Sec.\ \ref{continuum}). 
The generalized Kazakov potentials $V_s(x)$ where $s \in \mathopen{]}1,\infty\mathclose{]}$, allow for a  combinatorial interpretation 
which will be described in Sec.\ \ref{combinatorial} and the relation to $O(n)$ 
models on random triangulations is outlined in Sec.\ \ref{Onmodel}.  Finally Sec.\ \ref{discuss} summarizes our results.

\section{The Multicritical matrix model}\label{multicritical}

Let us consider the following $N\times N$ Hermitian matrix model
\beq\label{ja2}
Z = \int d M \, \e^{-N \tr V(M)},
\eeq
where 
\beq\label{ja3}
V(x) = \frac{1}{g} \tilde{V}(x), \quad 
\tilde{V}(x)=\sum_{n=1}^m v_{n} x^{2n}, \quad v_{1} = \frac{1}{2}.
\eeq
In the large-$N$ limit there is a one-cut solution, where the eigenvalues of $M$ 
condense in an interval $[-a,a]$ and the so-called resolvent (also called the disk amplitude)
\beq\label{ja3W}
W(z) = \frac{1}{N} \LA \tr \;\frac{1}{z-M} \RA  = \int_{-a}^a \d x \, \frac{\rho(x)}{z-x}
\eeq
is an analytic function of $z$ outside the cut. 

The large-$N$ solution for $W(z)$ is 
\be
W(z)=\int_0^a \frac{\d x}{\pi}\frac{xV'(x)}{(z^2-x^2)} 
\frac{\sqrt{z^2-a^2}}{\sqrt{a^2-x^2}},
\label{Weven}
\ee
where the condition $W(z) \to 1/z$ for $|z| \to \infty$ implies
\be
g(a^2)=\int_0^a \frac{\d x}{\pi}\frac{xV'(x)}{\sqrt{a^2-x^2}} \;=\;
 \sum_{n=1}^m  \frac{v_n \,a^{2n}}{B(n,\frac 12)},
 \quad\quad B(x,y)=\frac{\Gamma(x)\Gamma(y)}{\Gamma(x+y)}.
\label{bbcc}
\ee
 This fixes $g(a^2)$ as a polynomial of $a^2$ for a  polynomial potential. 
 
  We can rewrite the integral representation \rf{Weven}
 for $W(z)$ in terms of the function $g(a^2)$ instead of the potential $\tilde{V}(x)$.
 Let us introduce a special notation for this function 
 \be
\tU(a^2)=\int_0^{a} \frac{\d x}\pi \frac{x \tV'(x)}{\sqrt{a^2-x^2}}= 
\int_0^1 \frac{\d y}\pi \;\frac{G((ay)^2)}{\sqrt{1-y^2}},\quad G(x^2)=x \tV'(x)
\label{defU}
\ee
so the boundary equation~\rf{bbcc} reads
\be
\tU(a^2)=g(a^2).
\label{U=1a}
\ee
Then one has the following representation of $W(z)$
\be
g W(z)= \int_0^{a^2} \d A\, \frac{\tU'(A)}{\sqrt{z^2-A}} = \int_0^g 
\frac{\d \tilde{g}}{\sqrt{z^2 - a^2(\tilde{g})}}  .
\label{11a}
\ee
The proof is based on the identity
\be
\int_0^a \frac{\d x}{\pi}\frac{x^{2n}}{(z^2-x^2)} 
\frac{\sqrt{z^2-a^2}}{\sqrt{a^2-x^2}} =
\frac{1}{2 B(n,\frac 12)} \int_0^{a^2}
\d A\, \frac{A^{n-1}}{\sqrt{z^2-A}}.
\ee

 For a general potential defined by a convergent power series
 we have from \rf{bbcc} the relation
\be
\tilde{V}(x)=\sum_{n=1}^\infty v_n x^{2n}, \quad 
\tU(A) = \sum_{n=1}^\infty u_n A^n, \quad 
v_n= u_n\, B(n,\textstyle{\frac 12)}.
\label{Vgen}
\ee
Finally note that \rf{U=1a} and \rf{11a} lead immediately to the known equation
for the disk amplitude with one puncture:
\beq\label{ja15} 
\frac{ \d \;g W(z)}{\d g} = \frac{1}{\sqrt{ z^2 -a^2(g)}}.
\eeq

A so-called  $m^{\rm th}$ multicritical point
 of this matrix model is a point where 
 \beq\label{ja4}
 \frac{\d g(A)}{\d A}\Big|_{A=a^2_c} = \cdots = 
\frac{\d^{m-1} g(A)}{\d A^{m-1}}\Big|_{A=a^2_c} =0, \quad 
\frac{\d^{m} g(A)}{\d A^{m}}\Big|_{A=a^2_c} \neq 0.
 \eeq
 In order to satisfy this  requirement an even potential $\tV (x)$ has to be at least
 of order $2m$. If we restrict ourselves to potentials of this order\footnote{In Sec.\ 
 \ref{continuum} we consider more general polynomials.} $g(a^2)$ is fixed 
 to be of the form
 \beq\label{ja5}
 g(a^2)= g_* - c (a_c^2 -a^2)^m, \quad g_* = c \,a_c^{2m}, \quad  c = \frac{1}{4m a_c^{2m-2}}.
 \eeq
The value 
 $a_c^2 > 0$ can be chosen arbitrary, after which the coefficients $v_n$ are completely fixed. 

For convenience we choose $a_c =1$, i.e.\ 
 \beq\label{ja7}
 g(a^2)=  g_* -\frac{1}{4m} (1-a^2)^m,\quad g_* = \frac{1}{4m}.
 \eeq
From \rf{bbcc} and \rf{ja7} we obtain the coefficients $v_n(m)$ for the $m^{\rm th}$ 
multicritical Kazakov potential
\beq\label{ja6a}
v_n =  \frac{(-1)^{n-1}}{4m} 
 \begin{pmatrix} m \\ n \end{pmatrix}  B(n,{\textstyle \frac 12})= 
 \frac 14 \;\frac{\Gamma(n-m) \Gamma(\frac 12)}{\Gamma(1-m)\Gamma(n+\frac 12)\, n},
 \quad \quad n \leq m ,
 \eeq
 where the last equality should be understood as the limit where $m$ goes to 
 an integer. For future use we write the $m^{\rm th}$ Kazakov potential as  
 \beq\label{ja6}
 V_s(x) = \frac{1}{g(a^2)} \sum_{n=1}^m v_n(s) x^{2n},  \quad
 v_n(s ) = \frac 14 \;  
 \frac{\Gamma(n+\frac 12-s) 
 \Gamma(\frac 12)}{\Gamma(\frac 32-s)\Gamma(n+\frac 12)\, n},  \quad 
 s \to m + \frac 12.
 \eeq

\section{The generalized Kazakov potential}\label{general}

Let us now generalize the  potential \rf{ja6} by simply allowing $s$ in $v_n(s)$ to be 
a real number larger than 1/2.  We thus introduce
\be
\tV_s(x)=\sum_{n=1}^\infty  v_n(s)\, x^{2n} = 
 {}_3 F _2 \left( 1,1,\frac 32-s;2, \frac 32;  x^2 \right)\!\frac {x^2}2, \quad
v_n(s)=\frac 14 \, 
 \frac{\Gamma(n+\frac 12-s) 
 \Gamma(\frac 12)}{\Gamma(\frac 32-s)\Gamma(n+\frac 12)\, n},
\label{Vmult}
\ee 
where ${}_3 F _2$  is the generalized hypergeometric function.  
Formally, taking $s \to m+1/2$ the infinite sum
is automatically terminated at $n = m$, and the $m^{\rm th}$ multicritical Kazakov 
potential is reproduced. For $s \neq m+1/2$ the coefficients behave as $v_n(s) \sim n^{-s-1}$ for $n\to\infty$ and therefore $\tilde{V}_s(x)$ is a power series with radius of convergence equal to one. 

Given \rf{Vmult} we find
\be\label{jx2}
x \tV'_s(x)=
 {}_2F_1\left(1,\frac32-s,\frac32, {x^2}\right) x^2
={}_2F_1\left(1,s,\frac32, \frac{x^2}{x^2-1}\right)\frac{x^2}{1-x^2}
\ee
and further,  from \rf{Vgen}:
\beq\label{ja9}
U_s(A) = \frac{1- (1-A)^{s-1/2}}{4 (s-1/2)} , \quad 
U'_s(A) = \frac 14 (1-A)^{s-3/2}, 
\eeq
and 
\beq\label{ja9a}
g_s(a^2) = \frac{1- (1-a^2)^{s-1/2}}{4 (s-1/2)} ,\quad g'_s (a^2) = \frac 14 (1-a^2)^{s-3/2},
\quad g_s^* = \frac{1}{4 (s-1/2)},
\eeq
which is the most obvious generalization of  \rf{ja5}. 

If we formally apply \rf{11a} we find for the potential \rf{Vmult}
\bea
g W(z)&=&\frac 1{4}\int_0^{a^2} \d A\, \frac{(1-A)^{s-3/2} }{\sqrt{z^2-A}}~~ \label{Wmulti1}\\
&=& \frac{{}_2F_1 \left(1,\frac32-s,\frac{3}{2},{z^2}\right) z-
{}_2F_1\left(1, \frac32-s,\frac{3}{2},\frac{z^2-a^2}{1-a^2}\right)
(1-a^2)^{s-\frac{3}{2}} \sqrt{z^2-a^2}}{2}~~\label{Wmulti2}\\
&=&\frac{ {}_2F_1 \left(1,s,\frac{3}{2},\frac{z^2}{z^2-1}\right) z-
{}_2F_1\left(1,s,\frac{3}{2},\frac{z^2-a^2}{z^2-1}\right) 
(1-a^2)^{s-\frac{1}{2}} \sqrt{z^2-a^2}}{2 (1-z^2)}~~
\label{Wmulti3}\\
&=&\frac{ {}_2F_1 \left(1,s,\frac{1}{2}+s,\frac{1}{1-z^2}\right) z -
{}_2F_1\left(1,s,\frac{1}{2}+s,\frac{1-a^2}{1-z^2}\right) 
(1-a^2)^{s-\frac{1}{2}} \sqrt{z^2-a^2}}{4 (s-\frac 12 )(z^2-1)}~~~ ~ \label{Wmulti}
\eea
where the relation between $a$ and $g$ is given by \rf{ja9a},  i.e.
\beq\label{ja10}
a^2 = 1 - \left(1- \frac{g}{g_*}\right)^{\frac{1}{s-\frac{1}{2}}}, 
\quad g_* = \frac{1}{4 (s - 1/2)}.
\eeq
All the representations of $W(z)$ given above have their virtues as we will now 
describe.  

A standard representation of $g W(z)$ for an ordinary (even) polynomial $\tV(z)$  of degree $2n$ is 
\beq\label{jx1}
g W(z) = \frac 12 \left[\tV'(z) - M(z^2-a^2) \sqrt{z^2-a^2}\right], \quad M(x) = \sum_{k=1}^{n}
M_k\; x^{k-1},
\eeq
where  $M(x)$ is a  polynomial of degree $n-1$, 
uniquely fixed to cancel $\tV'(z)$ and to insure
that $W(z) \to 1/z$ for $ |z| \to \infty$.  In our case $\tV'(z)$ will have a cut along the real axis starting at $z^2=1$ as is clear from \rf{jx2}. Correspondingly $M(z^2-a^2)$ should thus have a similar cut and \rf{Wmulti2} is simply the representation \rf{jx1} and we have
\beq\label{jx3}
2g W(z)  - \tV'(z) = -M(z^2-a^2) \sqrt{z^2-a^2},\quad 
M(x) = (1-a^2)^{s-\frac{3}{2}}{}_2F_1\left(1, \frac32-s,\frac{3}{2},\frac{x}{1-a^2}\right),
\eeq
from which we can read off the coefficients $M_k$. 

\begin{figure}
\centering
\includegraphics[width=8cm]{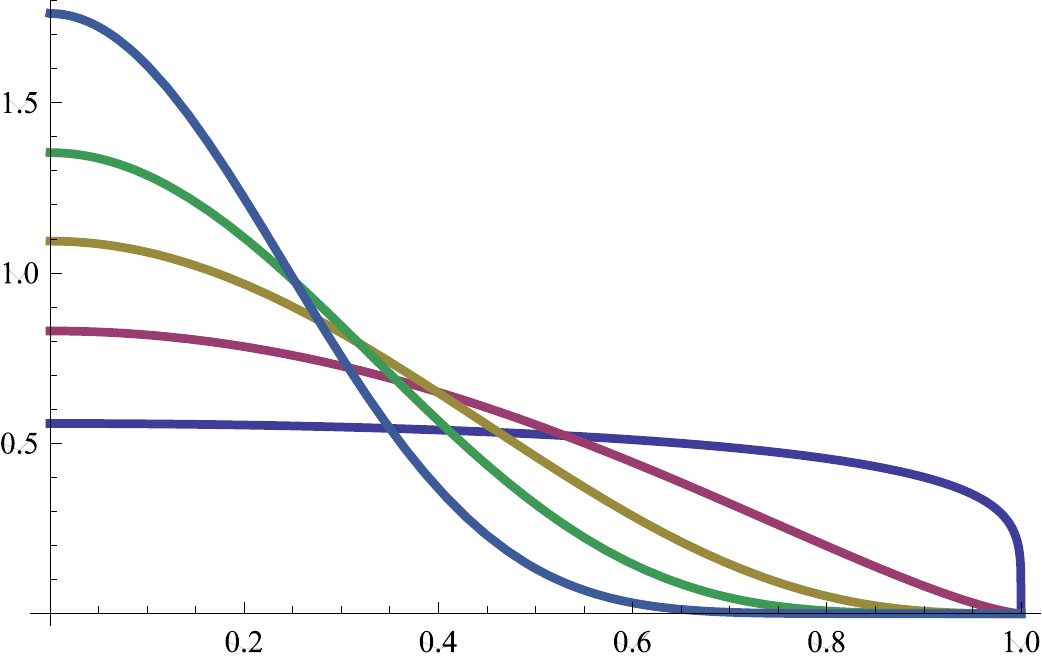}
\caption{Plot of $\rho(x)$  versus $x$ for $s=1.2$, $2.4$, $4$, $6$, $10$ from
bottom to top.}
\label{fi:rho}
\end{figure}

 The representation \rf{Wmulti3} is useful because the hypergeometric functions
 are analytic  along the cut $z \in [-a,a]$ of $W(z)$, $0 < a <1$, and 
 thus the discontinuity across the cut is entirely determined simply by the discontinuity
 of $\sqrt{z^2-a^2}$. 
From the very definition  \rf{ja3W} of $W(z)$ it follows that the density of 
eigenvalues, $\rho(x)$, is determined
by the discontinuity of $W(z)$ across the cut and we thus obtain:
\beq\label{ja16}
\rho(x) = \frac{ \lim_{\epsilon \to 0}(W(x +\i \epsilon) - W(x-\i \epsilon))}{2\pi \i} =  
\frac{(1-a^2)^{s-\frac{1}{2}} \sqrt{a^2-x^2}\,{}_2F_1\left(1, s,\frac{3}{2},\frac{a^2-x^2}{1-x^2}\right)}{2 \pi g(1-x^2)} .
\ee
This $\rho(x)$  is plotted as $a^2\to1$   (or $g\to g_*$) in Fig.~\ref{fi:rho} for several values of $s$. Up to normalization these plots correspond to $(1-x^2)^{s-1}$ since we can rewrite $\rho(x)$ as 
\beq\label{ja16a}
\rho(x) =
\frac{ \Gamma(s-\frac{1}{2})}{4\sqrt{\pi} g\Gamma(s)} \; (1-x^2)^{s-1}-
\frac{(g_*-g)}{\pi g} \left[ \frac{\sqrt{a^2\!-\!x^2}}{\!1-\!x^2}
 {}_2F_1\left(1, s,s\!+\!\frac{1}{2},\frac{1\!-\!a^2}{1\!-\!x^2}\right) \right] 
 \eeq
where the part in brackets is bounded for all $a \in [x,1]$ for fixed $x^2<1$.
Further, $\rho(x)$ is positive in $x \in \mathopen{]}-a,a\mathclose{[}$, vanishes at $x=\pm a$ 
and tends to the delta-function as $s\to\infty$. 

Finally the representation \rf{Wmulti} shows that $W(z)$ indeed has 
convergent power expansion  in $1/z$ for $|z|$ sufficiently large and 
using \rf{ja9} it follows that $W(z) \to 1/z$ for $|z| \to \infty$. For future reference 
we note that the transformation of the hypergeometric functions from \rf{Wmulti3}
to \rf{Wmulti} involves terms not seen in \rf{Wmulti}. More specifically one has 
\bea
 {}_2F_1 \left(1,s,\frac{3}{2},\frac{z^2}{z^2-1}\right)  & =& 
 {}_2F_1 \left(1,s,\frac{1}{2}+s,\frac{1}{1-z^2}\right) \;\frac{1}{2 (\frac 12 -s)} +\nonumber \\
&& +\i \frac{ (1-z^2)^{s} }{z} \frac{\sqrt{\pi} \Gamma(s-\frac 12)}{2\Gamma(s)} \label{jx4}
\eea
but the last term on the rhs of eq.\ \rf{jx4} cancels against an identical term coming from
the other hypergeometric function in \rf{Wmulti}.

Let us end this section by calculating the susceptibility exponent $\gamma_s$ associated with the matrix model with potential  \rf{Vmult}. We define the susceptibility as the second derivative of the free energy of the matrix model with respect to the coupling constant $g$:
\beq\label{ja20} 
F = \frac{1}{N^2} \log Z, \quad \chi = \left( g \frac{ \d }{\d g} \right)^2 F
\eeq
and $\gamma_s$ by
\beq\label{ja21}
\chi (g) = \chi_a (g) + c (g_* -g)^{-\gamma_s} + {\rm less~singular}.
\eeq
where $\chi_a(g)$ is analytic at $g_*$. Expanding  $\d (gW(z))/\d g$ in
inverse powers of $z$, any of the terms $c_n(g)/z^{2n+1}$, $n >1$, will
have $ (g_* -g)^{-\gamma_s}$ as the leading non-analytic term. From \rf{ja15}
and \rf{ja10} it follows immediately that the term is $(g_*-g)^{1/(s-1/2)}$ and therefore
\beq\label{ja22}
\gamma_s = - \frac{1}{s-\frac{1}{2}}.
\eeq

 For $s \in \mathopen{]}m-1/2,m+1/2\mathclose{[}$ with $m$ a positive integer our potential \rf{Vmult} has many of the 
characteristics of the $s = m+1/2$ multicritical potential: the first $m$ terms in the power series have  oscillating signs, starting out always with $x^2/2$.  The signs of terms  $x^{2n}$, $n \geq m$ are the same.                                                                                                                                                                                                                                                                                                                                                                                                                                                                                                                                                                                                                                                At the same time, moving $s$ towards $m+1/2$,  $\gamma_s$ changes
continuously towards the value $-1/m$ of the $m^{\rm th}$ multicritical model.
The range $s \in\mathopen{]}1/2,3/2\mathclose{]}$ is special. It starts out with $s = 3/2$, i.e. $m=1$ and 
thus $\tV(x) = x^2/2$, i.e.\ a trivial Gaussian potential and we have 
\beq\label{ja30} 
g W(z) = \frac{1}{2} (z - \sqrt{z^2-a^2}),\quad g = \frac{1}{4} \, a^2.
\eeq
$a^2$ is an analytic function of $g$, in accordance with the value $\gamma_s=-1$.
For $1/2 <s < 3/2$ all coefficients in the power series expansion of $\tV(x)$
are positive and the derivative $g'(a^2=1)$ is infinite rather than zero as 
for $s > 3/2$. For $s \to 1/2$, $g_* \to \infty$ while $\gamma_s \to -\infty$.

For the $m^{\rm th}$ multicritical potential it is well known that $\gamma = -1/m$ does
not correspond to the KPZ area susceptibility exponent $\gamma_A$ \cite{staudacher}. 
Rather, it is related to insertions of the primary operator with the most negative scaling
dimension, which in non-unitary conformal theories coupled to 2d gravity need not be the 
cosmological constant. In the multicritical models one obtains the KPZ exponent  by 
identifying the cosmological constant via the length of the boundary of the disk. 
One thus looks at  
\beq\label{jz1}
\LA W (2\ell)\RA  := \frac{1}{N} \LA\tr  \; M^{2\ell}\RA= 2\int_0^a \d x \, \rho(x)\, x^{2\ell}
\to a^{2l} \
\left[ \frac{(1-a^2)^{s-\frac 32} a^2 \Gamma(\ell + \frac 12)}{4 \sqrt{\pi}
g(a^2) \Gamma (\ell +2)} \right]
\eeq
where the average $\LA \cdot \RA$ is with respect to the partition function
\rf{ja2}. We are interested in the limit  
$\ell \to \infty$ where the integral  will be dominated by $x$ close
to the boundary $a$. One obtains the leading $\ell$ behavior 
\beq\label{jz2}
\LA W(2\ell)\RA \sim \exp (2\ell \, \log a + O (\log \ell)) = 
\exp \left( - \Big(1- \frac{g}{g_*}\Big)^{\frac{1}{s-1/2}} \, \ell + O (\log \ell)\right) ,
\eeq
where we have used \rf{ja10}. Thus we identify the dimensionless
boundary cosmological constant  $\mu_B$ and we introduce the dimensionless bulk cosmological constant $\mu \sim \mu_B^2$ as follows
\beq\label{jz3}
\mu_B \sim \left(1- \frac{g}{g_*}\right)^{\frac{1}{s-1/2}} =  
\left(1- \frac{g}{g_*}\right)^{-\gamma_s},
\quad \quad \mu \sim \left(1- \frac{g}{g_*}\right)^{-2\gamma_s}.
\eeq
From the definition \rf{ja20} we have 
\beq\label{jz4}
F(g)\Big|_{\rm singular} \sim (g_*-g)^{2-\gamma_s} \sim \mu^{2-\gamma_A}
\eeq
and we conclude that 
\beq\label{jz5}
\gamma_A = \frac 32 + \frac{1}{\gamma_s} = 2-s.
\eeq
If we assume that $\gamma_A$ is related to an underlying conformal 
field theory coupled to 2d quantum gravity, as is the case for the multicritical
points where $s = m+1/2$, we have from the standard KPZ relation
that the central charge of the matter fields related to $s$ is
\beq\label{jz6}
c(s) = 1 - 6 \frac{\gamma_A^2}{\gamma_A-1} = 1- 6 \frac{(s-2)^2}{s-1}.
\eeq
The same $c(s)$ corresponds to two different $\gamma_A$'s, related by
\beq\label{jz7} 
\gamma_A \to \gamma_A' = -\frac{\gamma_A}{1-\gamma_A},
\quad {\rm i.e.} \quad s \to s' = \frac{s}{s-1}.
\eeq
The two $\gamma_A$'s correspond to the two different solutions
to the KPZ relation \rf{jz6}. Usually the conformal field
theory associated with a given central charge $c$ is assigned 
a $\gamma(c)$ from the branch where $\gamma(c) \to -\infty$ for $c \to -\infty$.
However, the other branch also has an interpretation in terms of random 
surfaces and 2d quantum gravity \cite{durhuus,adj,klebanov}.

If we follow the above conjectures we are led to the following picture: $s = 2$ 
corresponds to $c =1$ ($ \gamma_A = 0$) where the two branches meet.
The region $s \in \mathopen]2, \infty\mathclose{[}$ corresponds to the ``physical'' branch of the KPZ equation
where $\gamma_A$ changes from 0 to $-\infty$. The other branch corresponds to 
$s' \in \mathopen{]}1,2\mathclose{[}$ and $\gamma_{A}' >0$, approaching 1 for $s' \to 1^+$. An interesting 
example is $s' = 3/2$ considered above. Formally it corresponds to the $m=1$ ``multicritical'' matrix model which is just the Gaussian matrix model 
with $W(z)$ given by \rf{ja30}. In KPZ context it can be viewed as the (2,1)
conformal field theory coupled to 2d gravity in the series of $(2,2m-1)$ conformal
field theories corresponding to the multicritical models, although it, contrary to 
the larger $m$ theories, is not a standard {\it minimal} conformal field theory. 
The KPZ assignment of central charge to this theory is $c = -2$ and 
the corresponding $\gamma_A = -1$. In fact we found $\gamma_s = -1$ above,
but according to \rf{jz5} the corresponding $\gamma_A' = 1/2$, in agreement 
with the fact that $W(z)$ in \rf{ja30} {\it is} the partition function for branched 
polymers which is known to have $\gamma = 1/2$. That branched polymers
play an important role in the interpretation of $\gamma_A'$ is the essence
of the work \cite{durhuus,adj,klebanov}. It also follows from \rf{jz7} 
that $s' =3/2 \to s=3$ and $s=3$ indeed gives $c = -2$ and $\gamma_A = -1$.
In Sec.\ \ref{combinatorial} we will see it is possible to give a  combinatorial
explanation of the relation between $s$ and $s'$ which is in agreement 
with the picture picture developed in \cite{durhuus,adj,klebanov}.

Clearly $s=1$ is special, being the limit where the assumed central charge 
$c(s) \to -\infty$ and $\gamma_{A'} \to 1$. The potential \rf{Vmult} is in this case
\beq\label{jz8}
\tV_{s=1}'(x) = \log \Big( \frac{1+x}{1-x}\Big)=2 \arctanh x,
\eeq 
and the corresponding disk function from \rf{Wmulti}
\beq\label{jz9}
W(z) = \frac{ {\rm arcsinh} \sqrt{ \frac{1}{z^2-1}} - 
\arctanh \sqrt{\frac{1-a^2}{z^2-a^2}}}{1-\sqrt{1-a^2}}.
\eeq
It is interesting that all potentials corresponding to integer $s >1$, i.e. non-negative 
integer $\gamma_A$, are simple modifications of \rf{jz8}. Similarly the corresponding
$W$s are  simple modifications of \rf{jz9}. These statements follow from Gauss' 
recursion relations for  hypergeometic functions.

\section{The Riemann-Hilbert method at work}\label{riemann-hilbert}

Above we assumed that one can use the standard large $N$ one-matrix model 
formula to obtain the disk function. Let us briefly discuss why the formula 
is still valid in certain cases where $V'(x)$ has  cuts and poles at the real axis. 
It represents a  simple generalization of the usual case of the one-matrix model with  polynomial $V'(x)$ which can still be treated by the  Riemann-Hilbert method. 

The large $N$ saddelpoint  of the matrix model is the principle value integral 
\be
 V'(x)=2  \dashint \d y  \frac{\rho(y)}{x-y},
\label{sp} 
\ee
which is valid when $x$ belongs to the support of 
the eigenvalue density $\rho$ which is assumed to avoid possible cuts and poles of $V'$.
We proceed in the usual way by introducing the analytic function 
\be
W(z) =  \int \d y  \frac{\rho(y)}{z-y},
\label{2}
\ee
and rewriting \eq{sp} 
at the real axis as
\be
\Im \left( W^2 -V' W \right)+\Im V'  \Re W=0 .
\label{3}
\ee
Usually, the term with $\Im V'$  is missing since $V'$ is real at the real axis, but 
we now have to include it since  $V'$ can have cuts located on the real axis.

Equation~\rf{3} on the real axis implies the following equation in the whole
complex plane:
\be
W^2(z) -V'(z) W (z)+ \int _{C_2} \frac{\d \omega}{2\pi \i}
\frac{V'(\omega) W(\omega)}{(z-\omega)}= Q(z) ,
\label{4}
\ee
where the contour $C_2$ 
encircles possible cuts and poles of $V'(\omega)$
on the real axis, but not  $z$ and not the cut(s) of $W(\omega)$.
$Q(z)$ is an entire function 
(a polynomial if  $V'$ is itself  a polynomial) and its role is to compensate nonnegative
powers of $z$ in the product $V'(z) W(z)$. The third term on the left-hand side of 
\eq{4} plays thus no role in determining $Q(z)$.

We can rewrite \eq{4} as
\be
W^2(z) - \int _{C_1} \frac{\d \omega}{2\pi \i}
\frac{V'(\omega) W(\omega)}{(z-\omega)}= 0 ,
\label{5}
\ee
where the contour $C_1$ encircles (anti-clockwise) the cut(s) of $W(\omega)$,
 but not  $z$ and possible cuts and poles of $V'(\omega)$.
We can prove the equivalence of Eqs.~\rf{4} and \rf{5} by deforming the contour 
$C_1$ in \eq{5} to $C_2$, which will give the third term on the left-hand side of \eq{4}. We get in addition the residual at $\omega=z$, which
accounts for the second term 
on the left-hand side of \eq{4}, and finally we get the contribution from $\omega=\infty$,
which is equal $Q(z)$.

Equation~\rf{5} is the usual loop equation of the one-matrix model 
at $N=\infty$ with the
potential $\tr\, V(M)$. Its standard derivation by an
infinitesimal shift of $M$ apparently works for all potentials,
including the ones with cuts on the real axis.
Correspondingly, \eq{5} results  in the usual 
formula for the one-cut solution
\be
W(z)=\int_a^b \frac{\d x}{2\pi}\frac{V'(x)}{(z-x)} \frac{\sqrt{(z-a)(z-b)}}{\sqrt{(x-a)(b-x)}},
\label{Wgen}
\ee
where the cut is  from $a$ to $b$. 
For an even potential $V(x)=V(-x)$, when the cut is from $-a$ to $+a$, it simplifies to
\rf{Weven}. The values of $a$ and $b$ are determined from the condition $W(z)\to1$ as
$z\to\infty$, which  for an even potential reduces to \rf{bbcc}.
Explicit formulas for a simplest non-even logarithmic potential are presented in
Appendix~A.

\section{Universality}\label{universality}

Let us recall the universality situation when the potential $V(x)$ is 
(an even) polynomial. Using a Wilsonian wording we have 
an infinite dimensional space of coupling constants, the coefficients in all polynomials
$V(x)$ and the $m^{\rm th}$ critical surface is characterized  by the condition 
\rf{ja4}. It has finite co-dimension $m-1$ and one can approach the surface
such that $m-1$ parameters survive in the ``continuum'' limit (see \cite{book} for a 
review). The Kazakov  potential \rf{ja6} is a particular simple choice of polynomial
which only depends on one parameter, $g$. We would like to understand  the universality 
situation for the new critical points defined by the generalized Kazakov potentials
$V_s(x) = \frac 1g \tV_s(x)$.

Clearly the new critical behavior is related to the tail $v_n \sim n^{-1-s}$
in $\tV(x)$. Let us choose another potential with the same tail but depending
on two parameters, $g$ and $c$, rather than the single $g$  in $V_s(x)$, 
\be
 \hat{V}(x)=\frac 1g \left[\frac{x^2}2 (1+ c) -\frac c{2} {\rm Li}_{1+s}\left( x^2\right)
\right]=\frac 1g \left[
\frac{x^2}2-\frac c{2}\sum_{n=2}^\infty \frac{x^{2n}}{n^{1+s}}\right],
\label{VV5}
\ee
where $ {\rm Li}_{1+s}$ is the polylogarithm.
This potential is rather general. In particular, we can  get a quartic
potential  from \rf{VV5} in the limit $c\to \infty$, $g\sim 1/c$.

The boundary equation \rf{U=1a} now reads
\be
g(a^2) =  \frac14 \left[a^2 (1+c) - c F_s(a^2) \right],
\label{1ga}
\ee
where the function $F_s(A)$ (trivally related to $\tU(A)$) is defined by
\beq\label{defF}
F_s(A) = \int_0^{\sqrt{A}} \frac{\d x}{\pi} 
\frac{ {\rm Li}_{s}( x^2)}{\sqrt{A-x^2}}
=\frac2{\Gamma(s)}
\int_0^\infty \d \tau \,\tau^{s-1}\left(  \frac1{\sqrt{1-A \e^{-\tau}}}-1 \right) .
\eeq
It has the following expansion (see also \rf{Vgen})
\beq
F_s(A)= \sum_{n=1}^\infty \frac{2 A^n}{B(\frac 12 ,n)\,n^{s+1}}
=A+\frac{3}{2^{2+s}}A^2+ \ldots .
\label{defFaa}
\eeq

Using the properties of $F_s(A)$ listed in Appendix~B,  one can 
analyse the function $g(a^2)$. It is an analytic function of $a^2$ for $0 \leq a < 1$ and the behavior close to $a^2=1$ is as follows:
\beq\label{ja40}
g(a^2) = f(1-a^2)  - \frac{2c}{\sqrt{\pi}} (1-a^2)^{s-1/2} \Big(1+O((1-a^2)\Big),
\quad s>3/2,
\eeq
where $f(x)$ can be expanded to order $[s-1/2]$:
\beq\label{ja41}
f(x) = g_s^* +g'(1) \; x + O(x^2), \quad g_s^* = \frac{1}{4} (1+c - c F_{s}(1)),
\quad g'(1)= -\frac{1}{4} (1+c - c F_{s-1}(1)).
\eeq
The function $g(a^2)$ starts out as an increasing function of $a^2$. By increasing $a$,
eventually $a$ might become a non-analytic function of $g$. This happens either
at the first $a$ where $g'(a) = 0$ or, if $g'(a) > 0$ for all $a$, at $a=1$, the radius of 
convergence for $g(a^2)$.  In the former case we have an $a_c < 1$ where
$g'(a_c) = 0$ and a corresponding critical value of $g$, $g_c = g(a_c)$.
For our choice of the potential (depending only on $g,c$) one can show that $g''(a) \neq 0$ for all values of $a <1$.

\begin{figure}
\centering
\includegraphics[width=8cm]{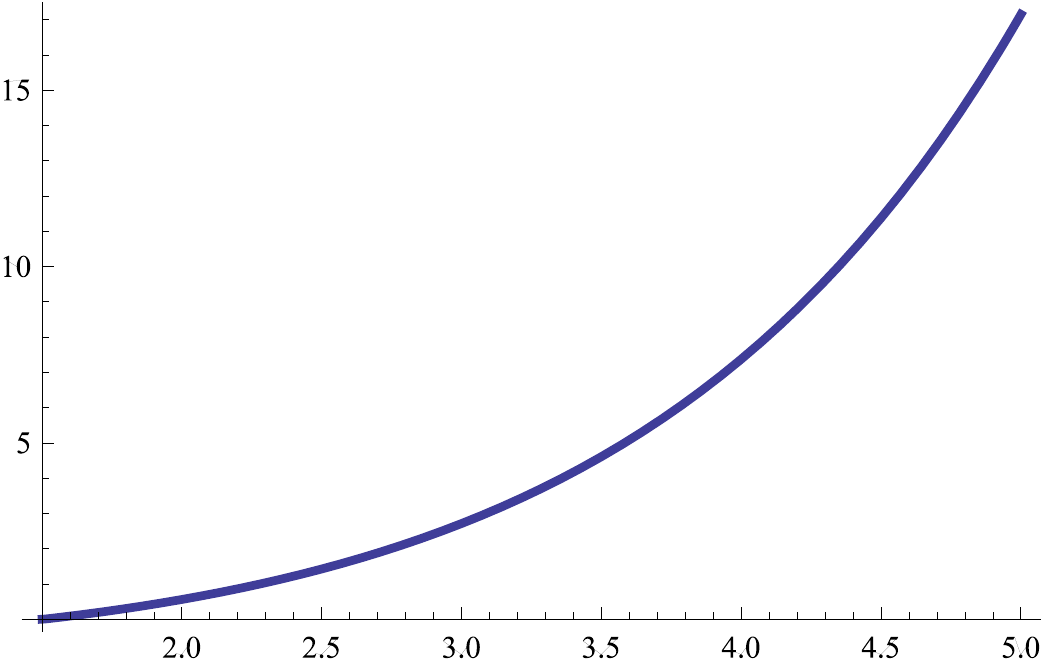}
\caption{Value of $c_*$ versus $s$ above which the usual 2d gravity 
scaling limit is realized for the polylog potential \rf{VV5}.}
\label{fi:2}
\end{figure}

In a neighborhood of $a_c$ we can therefore write
\beq\label{ja42}
g(a^2) = g(a_c^2) - k^2 (a_c^2-a^2)^2.
\eeq
We thus conclude that the leading non-analytic behavior of $a$ as a function
of $g$ is  $(a_c^2 -a^2)^{1/2}$, i.e. we have 
the standard situation with $\gamma_s = -1/2$,
corresponding to the $m=2$ Kazakov potential.  Whether or not this situation is realized
depends on the value of $c$.  We have 
\be
a^2 \frac {\d}{\d a^2} g =\frac 14 \left[a^2 (1+c ) - c F_{s-1}(a^2),
\right]
\label{d1ga}
\ee
and thus the following equation for the  value $c_*(s)$ separating the
two situations:
\beq\label{ja43}
 a^2 \frac {\d}{\d a^2} g\Big|_{a=1} = 0, \quad {\rm i.e.}\quad 
c_*(s)=\frac 1{F_{s-1}(1)-1}~~~(\geq 0).
\eeq 
This $c_*(s)$ is  positive for $s>3/2$, because then $1<F_{s-1}(1)<\infty $ and 
increases rapidly with $s$ as is depicted in Fig.~\ref{fi:2}.

Let us first discuss the situation for $s \in \mathopen{]}3/2,5/2\mathopen{[}$. 
For a given $s$ in this interval and a given $c \leq c_*(s)$ the critical point  is thus 
$g_s^*$ corresponding to $a_c = 1$ and the relation between $a$ and $g$ close
to $a_c$ is determined by \rf{ja40} and \rf{ja41}. For fixed $c < c_*(s)$ the  analytic 
term from $f(1-a^2)$ will dominate over the non-analytic term $(1-a^2)^{s-1/2}$ 
and we  have formally the situation corresponding to $\gamma = -1$. 
However, precisely for $c= c_*(s)$ this term will by definition vanish
and we obtain  from \rf{ja40}
\beq\label{ja44} 
g(a^2) = g_s^* - \frac{2c_*(s)}{\sqrt{\pi}} (1-a^2)^{s-1/2},\quad s \in\mathopen{]}3/2,5/2\mathopen{[},
\eeq
i.e.\ precisely the same scaling relation as for the generalized Kazakov potential,
and thus also $\gamma_s = 1/(1/2-s)$. If $c > c_*(s)$ we have $\gamma_s = -1/2$,
but for $c\to c_*(s)$ \rf{ja44} will take over since the term non-analytic in $(1-a^2)$ 
will dominate over the contribution \rf{ja42} when $a_c \to 1$. In the limit $s \to 5/2$
they will agree and give $\gamma_{5/2} = -1/2$.

 If we consider $s \in \mathopen{]}5/2,7/2\mathclose{[}$ we still have the same the curve $c_*(s)$, and results
 identical to those for $s \in\mathopen{]}3/2,5/2\mathopen{[}$ if $c \neq c_*(s)$. For $c= c_*(s)$ the term 
 in $f(1-a^2)$  linear in $(1-a^2)$ will still cancel, but the term proportional
 to $(1-a^2)^2$ will be dominant compared to $(1-a^2)^{s-1/2}$.  
 Only if we can cancel the analytic  $(1-a^2)^2$ term
 will we obtain a scaling like \rf{ja44} also for $s \in \mathopen{]}5/2,7/2\mathclose{[}$. To obtain such a 
 cancellation we need one further adjustable coupling constant apart from 
 $g$ and $c$. 

\begin{figure}
\centering
\includegraphics[width=8cm]{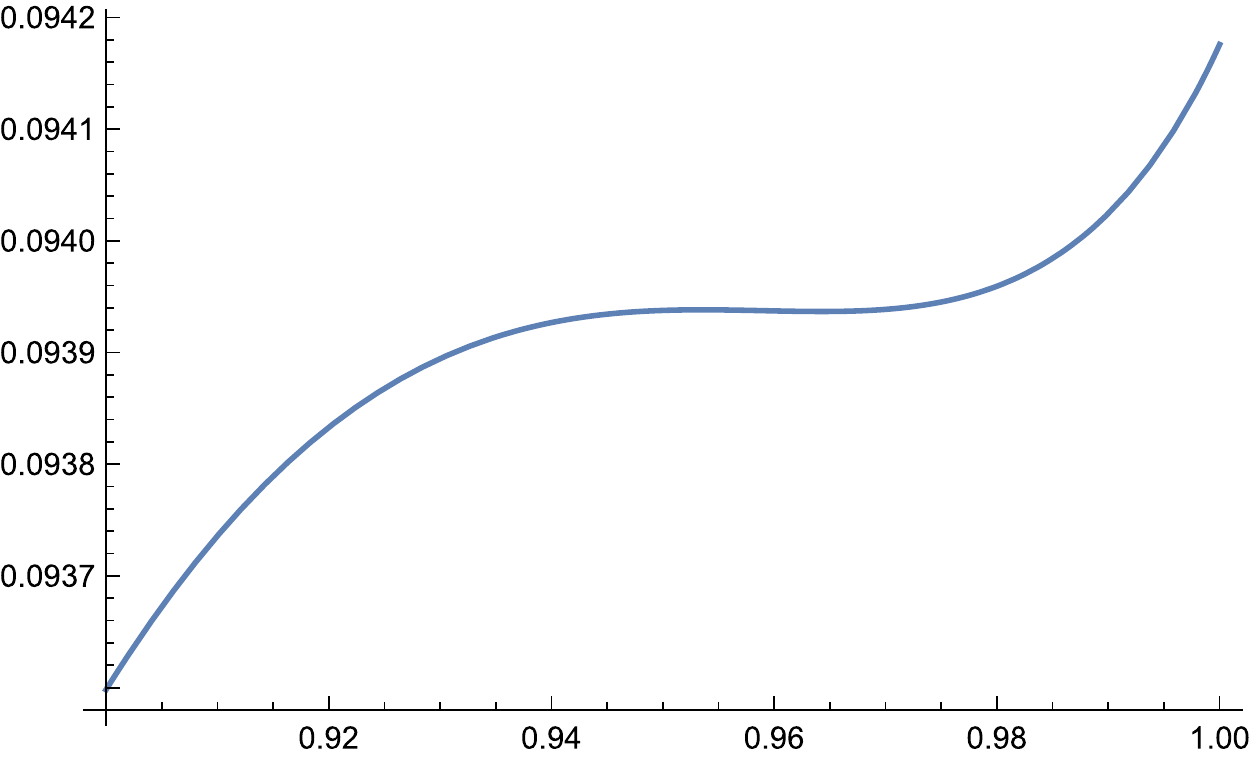}
\caption{Plot of $g(a^2)$ for $s=3$, $c=-3$ and $v_2 = -0.3433$ where 
$g'(a_c)=g''(a_c)=0$ for an $a_c <1$.}
\label{figmulti}
\end{figure}

\begin{figure}
 \centering
 \includegraphics[width=.45\linewidth]{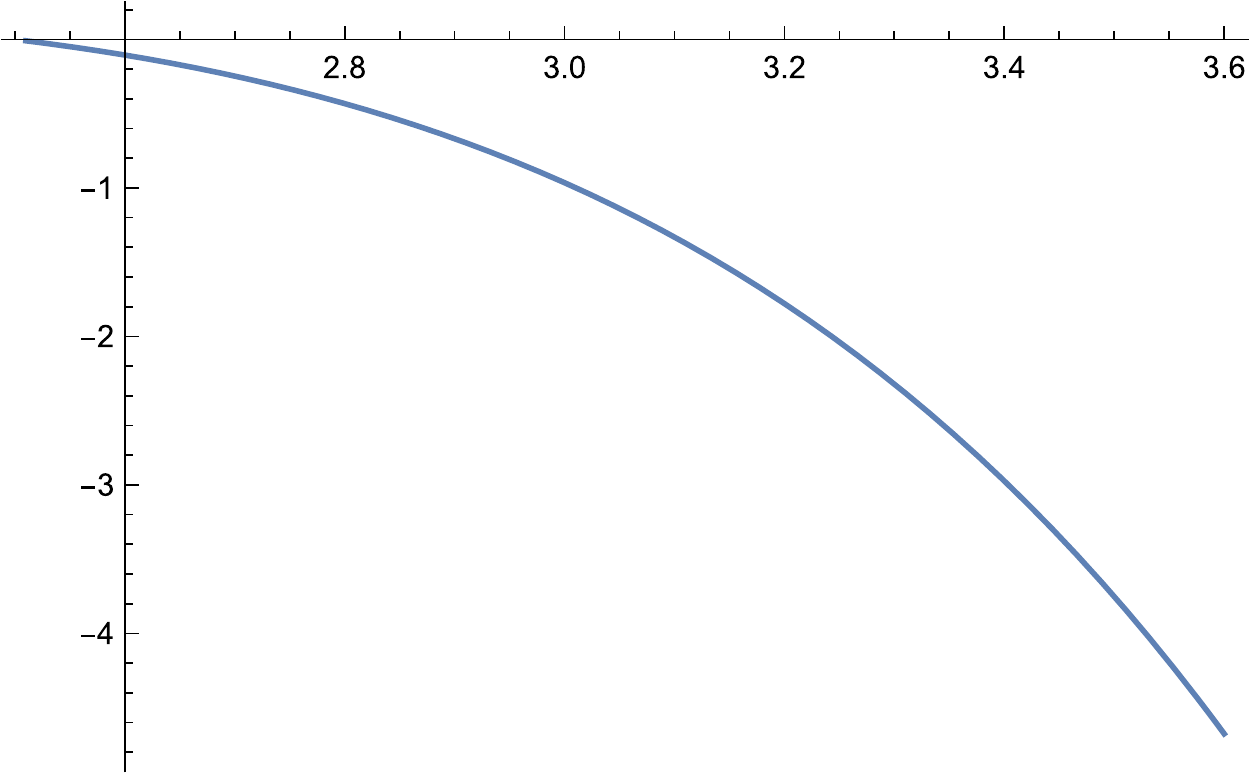}\hspace{.1\linewidth}\includegraphics[width=.45\linewidth]{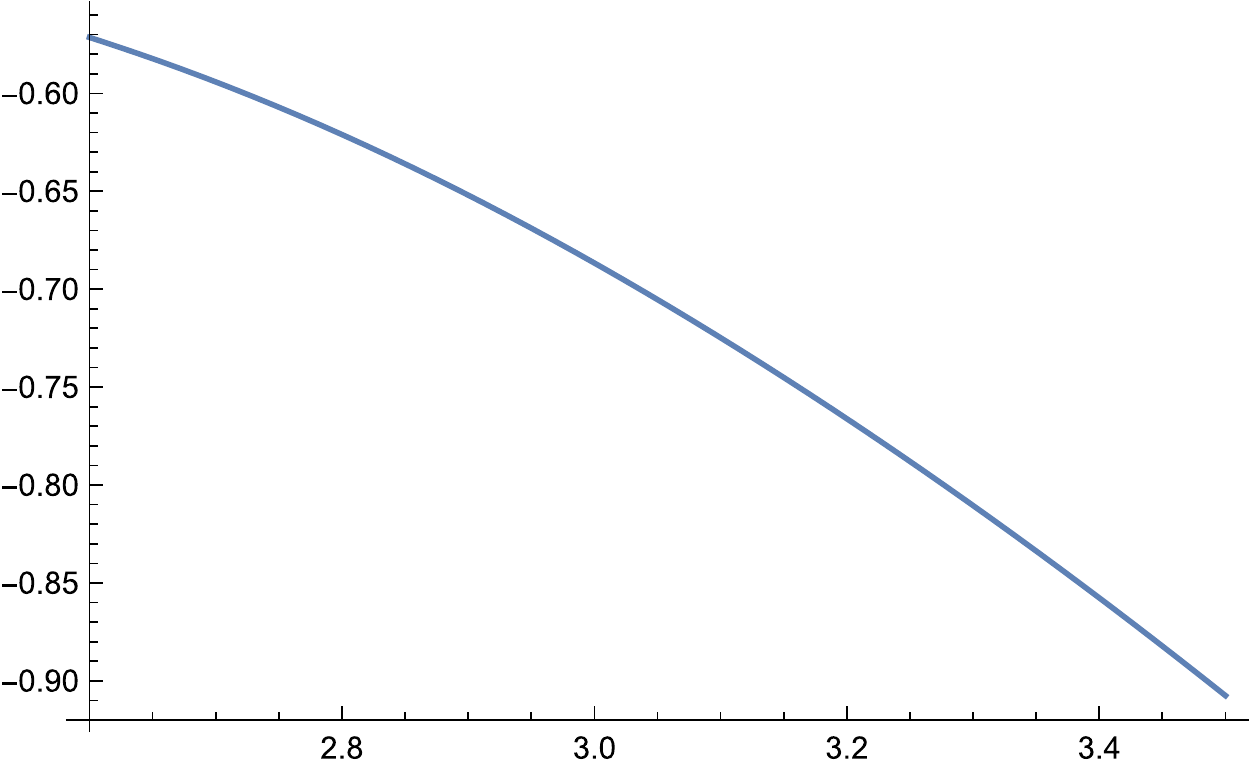} 

\caption{The left figure plots $c_*(s)$ from \rf{ja45} for $s > 5/2$. $c_*(s) \to 0$ for 
 $s \to 5/2$ since $F_{s-2}(1) \to \infty$ for  $s \to 5/2$. The right figure plots $v_2(s)$
 given by eq.\ \rf{ja46}.}
\label{figj1}
\end{figure}

There are many ways to introduce such a coupling constant, but 
 maybe the simplest is to add a term $v_2 x^4$ to the potential \rf{VV5}. With
 this new coupling constant at our disposal we can always find a point $a_c < 1$
 such that $g'(a_c^2)=0$. We can also try to find a point 
$a_c$ where  not only $g'(a^2_c) =0$ but also $g''(a^2_c)=0$, precisely as for the $m=3$ multicritical matrix model. In Fig.\ \ref{figmulti} we show such a situation.
Whether or not this is possible depends again on $c$ and corresponding
 to \eq{ja43} one obtains
 \beq\label{ja45}
 \Big(a^2 \frac {\d}{\d a^2}\Big)^n g\Big|_{a=1} = 0,~~n=1,2,
  \quad {\rm i.e.}\quad 
c_*(s)=\frac 1{2F_{s-1}(1)-F_{s-2}(1)-1} ~~~(\leq 0).
\eeq 
 and the corresponding value of $v_2(s)$ is 
 \beq\label{ja46}
 v_2(s)=\frac{4}{3} u_2(s) = \frac{2 c_*(s)}{3} (F_{s-2}(1)-F_{s-1}(1))~~~(< 0).
 \eeq
 We show $c_*(s)$ and $v_2(s)$ in Fig.\ \ref{figj1}. Note that they are both negative.
 For $c <c_*(s)$ we can approach $c_*(s)$ by changing $c$ while satisfying
 $g'(a_c)=g''(a_c)=0$, where $a_c(c) < 1$ and $a_c(c) \to 1$ for $c \to c_*(s)$.
 The condition $g'(a_c)=g''(a_c)=0$ determines $v_2$ uniquely for fixed $c$.
 For $c>c_*(s)$ one can approach $c_*(s)$ in such a way that $g'(a_c) =0$.
 This does not fix $v_2$ and the corresponding $a_c$, but by demanding that 
 $v_2 \to v_2(s)$ given by \rf{ja46}  we have by construction that $a_c \to 1$ and $g''(a_c) \to 0$ for $c \to c_*(s)$. For $s \in \mathopen{]}5/2,7/2\mathclose{[}$
 we thus have a situation completely analogous to $s \in\mathopen{]}3/2,5/2\mathopen{[}$, except that the 
 multicriticality while approaching $c_*(s)$ has changed from $m =1$ and 2 to 
 $m = 2$ and 3. At $c_*(s)$ we have $\gamma_s = -1/(s-1/2)$ and the potential $\hV(x)$
 is qualitatively the same as the generalized Kazakov potential $V_s(x)$ in the 
 same range of $s$.

 The generalization to higher values of $s$ is straight forward. For 
 $s \in \mathopen{]}m-1/2,m+1/2\mathclose{[}$ we allow deformations of $\hV$ involving $v_2,\ldots,v_{m-1}$.  We can define a critical $c_*(s)$ and approach it 
 from the two sides via $m-1$ and $m$ critical points by changing $c$, and
 the potential $\hV(x)$ at $c = c_*(s)$ will be qualitatively the same as $V_s(x)$.
 We have thus seen that the new scaling limits for $ s > 3/2$ are universal
 in the same way as the standard multicritical points of the one-matrix model 
 which correspond to $s=m+1/2$. 

Let us finally consider the region $s \in \mathopen{]}1/2,3/2\mathopen{[}$. 
 For $s$ in this region we have 
 \beq\label{ja47}
 g(a^2) = g^*_s - \frac{2c}{\sqrt{\pi}} (1-a^2)^{s-1/2} + O(1-a^2) .
 \eeq 
 Thus $a_c=1$ and $\gamma_s = -1/(s-1/2)$ and no fine tuning of $c$ is needed
 (except if one insists on $g^*_s$ positive one has to choose $c$ negative).
 The range of $\gamma_s$ is from $-1$ to $-\infty$, i.e.\  outside the range 
 of the original Kazakov range of $\gamma(m) = -1/m$  with integer $m$.  
 
 \section{The continuum limit}\label{continuum}
 
 For the (even) matrix models the scaling limit is usually performed 
 by the following assignment 
 \beq\label{ja50}
 a^2 \to a_c^2 -\sqrt{\Lambda} \epsilon,\quad z^2 = a_c^2 + P \epsilon.
 \eeq
 In our case $a_c =1$.  For most of the ``observables'' considered for 
 matrix models, this scaling is straight forward and unproblematic. As examples
 we have for the disk amplitude with one puncture, $\d (gW(z))/\d g$, that  
 \beq\label{ja51}
 \frac{\d \;gW(z)}{\d g} = \frac{1}{\sqrt{z^2-a^2}} \to \frac{1}{\sqrt{\epsilon} } \; 
 \frac{1}{\sqrt{P+\sqrt{\Lambda}}}
 \eeq
 and for the universal two-loop function (which can be derived for our more 
 general potentials precisely as for the ordinary polynomial potentials \cite{am,ajm}):
 \bea\label{ja52}
 W(z_1,z_2) &=& \frac{a^4}{2 (z_2\sqrt{z_1^2-a^2}+z_1 \sqrt{z_2^2-a^2})^2} 
 \frac{1}{ \sqrt{z_1^2-a^2}\sqrt{z_2^2-a^2}} \\
 &\to& \frac{1}{\epsilon^2} \;\;
 \frac{1}{\Big(\sqrt{P_1+\sqrt{\Lambda}} + \sqrt{P_2+\sqrt{\Lambda}}\Big)^2}\;\;
 \frac{1}{\sqrt{P_1+\sqrt{\Lambda}}\sqrt{P_2+\sqrt{\Lambda}}}.
 \eea
 The same is true for any higher loop functions. Approaching the $m^{\rm th}$ 
 multicritical point for ordinary matrix models one obtains
 \beq\label{ja53}
 W(z_1,\ldots,z_b) \to \frac{1}{\epsilon^{(b-2)m +\frac 32 b -1}} \; 
 W^{\rm cont}(P_1, \ldots, P_b; \Lambda),\quad b >2,
 \eeq 
 where $W^{\rm cont}(P_1, \ldots, P_b;\Lambda)$ denotes the {\it continuum}\/ $b$-loop 
 function\footnote{For the continuum two-loop function defined by eq.\ \rf{ja52}
 one often makes a subtraction which is irrelevant for our discussion, see 
 e.g. \cite{book}.}. The one (natural)  difference in our more general case will be that
 in the divergent pre-factor $m$ is replaced by $s-1/2$.

 The so-called continuum limit of the disk amplitude requires a more detailed 
discussion since it contains  a non-scaling part. If we use the representation
\rf{jx1} the potential term $\tV'(z)$ will not scale when
using the prescription \rf{ja50}. On the other hand the rest of the expression
will scale, as is clear from \rf{jx1} for a polynomial potential and from \rf{jx3}
for the generalized Kazakov potential. However the rhs of \rf{jx3} does not 
fall off as a function of the continuum $P$ for $|P| \to \infty$ the way one 
requires for the continuum disk-amplitude $W(P)$. One 
cures this by introducing a ``continuum'' potential $V_{\rm cont}(P)$
which is determined by the  requirement\footnote{It is often required 
that  the power series of $W_{\rm cont}(P)$ starts with the term
$P^{-3/2}$, i.e. one includes the first term $1/\sqrt{P}$ in $V'_{\rm cont}(P)$.} that  
$W_{\rm cont}(P)$ has a power expansion in $P^{-n-\frac 12}$, $n\geq 0$, for 
$P \to \infty$. We thus write%
\footnote{Sometimes a factor of 2 is inserted on the rhs of this formula to emphasize 
a doubling of continuum degrees of freedom for an even potential owing to
the symmetry $z\to -z$.}
\be
\left(g W(z)-\frac{\tV'(z)}2 \right)
={\epsilon^{s-1}} \left(W_{\rm cont}(P)-\frac{ \tV'_{\rm cont}(P)}2 \right). 
\label{43}
\ee
That the scaling factor is $\epsilon^{s-1}$ follows immediately from \rf{jx3}.
When $s=m+1/2$ it reduces to the ordinary scaling factor for the  ordinary $m^{\rm th}$ multicritical matrix model. Equations \rf{Wmulti}, \rf{jx3} and \rf{jx4} and the remarks 
surrounding \rf{jx4} allow us immediately to substitute the continuum limit 
\rf{ja50} and we obtain 
\beq\label{ja54}
W_{\rm cont} (P) = -g_s^* \,(\sqrt{\Lambda})^{s-\frac 12} \; \frac{ \sqrt{ P + \sqrt{\Lambda}}}{P}\;
{}_2F_1 \Big(1,s, \frac 12 + s; - \frac{ \sqrt{\Lambda}}{ P}\Big),
\eeq
\beq\label{ja55}
\tV'_{\rm cont} (P) = \i \frac{\sqrt{\pi} \Gamma(s-\frac 12)}{ 2\Gamma(s)} \;
(-P)^{s-1} .
\eeq
This $W_{\rm cont}(P)$ can indeed be expanded in powers  $1/P^{n+\frac 12}$ and the series is absolutely convergent for $|P| > \sqL$ and from the integral representation
of hypergeometric functions it follows that it is analytic for positive $P$.
 It has   a cut for negative $P$ starting at $P = - \sqrt{\Lambda}$,
 coming from $\sqrt{ P + \sqrt{\Lambda}}$. Like for the ordinary matrix models,
this cut is the scaled version of the original cut $[-a,a]$ in $z$. The potential 
$ \tV'_{\rm cont} (P)$ in \rf{ja55}
has a cut along the positive $P$ axis. This is the scaled 
version of the original cut of $\tV'(z)$ starting at $z=1$. 

If $s = m+1/2$ it is instructive to rederive the standard continuum 
results for the $m^{\rm th}$ model directly from \rf{jx3}. Using \rf{jx3} we obtain
\beq\label{jzz1}
\Big(g W(z)-\frac{\tV'(z)}{2}\Big)
=-\frac 12 \epsilon^{m-1/2} \Big[(\sqL)^{m-1}  {}_2F_1 \Big(1,1-m, \frac 32; 1 +
\frac{P}{ \sqrt{\Lambda}}\Big)\Big]\;\sqrt{P+ \sqL} , 
\eeq
where the expression part in square brackets is a polynomial in $P$ of order $m-1$, which can be written as 
\beq\label{jzz2}
 -g^* P^{m-1} 
\sum_{k=0}^{m-1} (-1)^{m-k}\frac{c_k}{c_m} \Big(\frac{\sqL}{P}\Big)^k,\quad 
c_k = \frac{\Gamma (k+ \frac 12)}{\Gamma (\frac 12)\Gamma (k+1)},
\eeq
$c_k$ being the coefficients in the Taylor expansion of $1/\sqrt{1-x}$.
This implies that except for $P^{m-1/2}$ all positive powers of $P$ will cancel 
on the rhs of eq.\ \rf{jzz1} and we obtain 
\beq\label{jzz3}
\Big(g W(z)-\frac{\tV'(z)}{2}\Big) = 
\frac{(-1)^m \Gamma (\frac 12) \Gamma (m) }{4 \Gamma (m + \frac 12)} P^{m-\frac 12}  -
g^*\frac{\Lambda^{m/2}}{\sqrt{P}} + O( P^{-3/2}),
\eeq
i.e.\ precisely the representation \rf{43}-\rf{ja55}.

 Let us briefly discuss the perturbation away from one of the 
 generalized multicitical points. One convenient 
 way to characterize the deformation away from the ordinary 
 $m^{\rm th}$ multicritical point is to use the so-called moments
 $M_k$ \cite{am,ajm,ackm}. They are defined by \footnote{An equivalent definition is
 $$
 M_k = \oint_C \frac{\d x}{2\pi \,i} \; \frac{x \tV(x)}{(x^2-a^2)^{k+1/2}},
 $$
 where the contour $C$ encircles to cut of $W(z)$ but not any poles or cuts of $V'(x)$.} 
 \beq\label{js1}
 M_k(a^2,v_n) = \frac{2}{k! c_k} \left(\frac{\partial}{\partial a^2}\right)^k \tU(a^2,v_n).
 \eeq
 In \rf{js1} we view $\tU$ and $M_k$ 
 as functions of $a^2$ and the coupling 
 constants $v_n$. For a given choice of coupling constants $v_n$  and $g$
 the position or the cut, i.e.\ the determination of $a$ as a function of $v_n$ and 
 $g$ will then finally be determined by  
 \rf{U=1a}. The coupling constants $v_n^c$ and $g_*$ correspond to an  $m^{\rm th}$ multicritical point if the  corresponding value $a=a_c$ 
 is such that  $M_k(a_c,v_n^c) = 0$, $k=1,\ldots,m-1$, $M_m(a_c,v_n^c) \neq 0$.
 In the case of the Kazakov potential we have chosen a particular simple
 way to move away from the critical point, namely by keeping the $v_n=v_n^c$
 and only changing $g$ and that case we had explicitly
 \beq\label{ju1}
 M_k(a^2) \propto  (1-a^2)^{m-k},\quad 0 < k \leq m,\quad M_k = 0, \quad k > m.
 \eeq
 For the generalized Kazakov potential this is changed to
 \beq\label{ju2}
 M_k(a^2) \propto  (1-a^2)^{s- \frac 12 -k},\quad k> 0,
 \eeq
the difference being that now infinitely many moments are different from zero.

 For the $m^{th}$ multicritical model a general deformation away from the
 multicritical point could be described as a change of coupling constants 
 away from the critical values such that 
\beq\label{ju3}
M_k =\mu_k \epsilon^{m-k},\quad 1\leq k \leq m,,\quad a^2 = a^2_c - \sqL \epsilon
\eeq
where $\mu_k$ and $\sqL$ are kept fixed when the  coupling constants 
change towards their critical values. As shown in \cite{ackm} all multiloop functions
can in the continuum limit be expressed as functions of $\mu_k$'s, $\sqL$ and 
the variables $P_1, \ldots, P_b$. The obvious generalization to a deformation
around the generalized Kazakov potential is to assume that
 \beq\label{ju4}
M_k =\mu_k \epsilon^{s-1/2-k},\quad 1\leq k < \infty,\quad a^2 = a^2_c - \sqL \epsilon
\eeq 
and that the  $\mu_k$'s and $\Lambda$ are kept fixed when the coupling 
constants flow towards their critical values.
With such a behavior all formulas for multiloop functions derived for 
the deformation around an arbitrary $m^{\rm th}$ model will remain valid
of any choice of $s$. For an arbitrary $s > 1/2$ it is possible to define so-called
continuous times $T_k$, related to  the $\mu_k$'s,  and to study the
so-called \emph{KdV} flow equations in terms of the $T_k$'s. Details of this will appear in a forthcoming paper \cite{to-appear}.

\section{Combinatorial interpretation}\label{combinatorial}

\begin{figure}
\centering
\includegraphics[width=.9\linewidth]{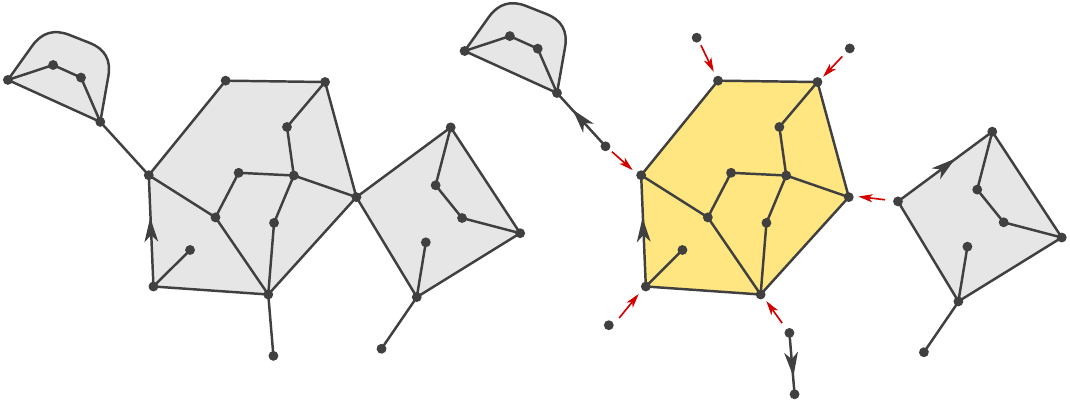}
\caption{A general rooted planar map with non-simple boundary (left) can be obtained uniquely by gluing rooted planar maps to each boundary vertex of a planar map with simple boundary (right).}
\label{fi:boundary}
\end{figure}

To better understand the duality $s\to \frac{s}{s-1}$ discussed in Sec.\,\ref{general} let us have a look at the combinatorial interpretation of the matrix model in terms of planar maps, i.e. graphs embedded in the plane modulo orientation-preserving homeomorphisms.
The boundary of a planar map $m$ is the contour of its ``outer face'', and we assume that $m$ has a distinguished oriented edge on the boundary, which is called the root edge.
We denote by $\mathcal{M}^{(l)}$, $l\geq 1$, the set of all such rooted planar maps that are bipartite, i.e. having all faces of even degree, and have boundary length $2l$.
By convention we let $\mathcal{M}^{(0)}$ contain a single map consisting of just a vertex.
If we write $q_n := \delta_{n,1} - 2n\, v_n$ for $n\geq 1$ then the disk amplitude $W(z)$ for $z^2 \geq a^2(g)$ and $g \leq g_*$ can be expressed as the convergent sum
\begin{equation}\label{diskcomb}
W(z) = \sum_{l=0}^\infty z^{-2l-1} \sum_{m\in\mathcal{M}^{(l)}} g^{\#\text{Vertices}(m)-1} \prod_{f\in \text{Faces}(m)}q_{\deg(f)/2}. 
\end{equation} 
One should notice that $\mathcal{M}^{(l)}$ contains planar maps with a boundary of the most general ``non-simple'' form, meaning that it may have pinch points in the sense that vertices appear multiple times in the boundary contour (see figure \ref{fi:boundary} for an example).
As we will see shortly, if $s \leq 2$ dropping the contribution of planar maps with non-simple boundaries from \eqref{diskcomb} has a non-trivial effect on the scaling properties of the disk amplitude.

We denote by $\hat{\mathcal{M}}^{(l)} \subset \mathcal{M}^{(l)}$ the planar maps with a ``simple'' boundary, meaning that all vertices in the boundary contour are unique, and define the simple disk amplitude $\hat{W}(x)$ for $x^2$ sufficiently small by
\begin{equation}
\hat{W}(x) := \sum_{l=0}^\infty x^{2l} \sum_{m\in\hat{\mathcal{M}}^{(l)}} g^{\#\text{Vertices}(m)} \prod_{f\in \text{Faces}(m)}q_{\deg(f)/2}.
\end{equation}
From the dual point of view $W(z)$ and $\hat{W}(x)$ can be interpreted respectively as the disconnected and connected planar Green functions, and it has long been recognized that they satisfy a simple relation \cite{BIPZ,GG95}.
Indeed, since a planar map $m$ with non-simple boundary contains a unique submap with simple boundary sharing the same root edge (see figure \ref{fi:boundary}), one easily observes that
\begin{equation}
W(z) = \frac{1}{gz}\sum_{l=0}^\infty (W(z))^{2l}\sum_{m\in\hat{\mathcal{M}}^{(l)}} g^{\#\text{Vertices}(m)} \prod_{f\in \text{Faces}(m)}q_{\deg(f)/2}=\frac{1}{g z} \hat{W}(W(z)).
\end{equation}
This implies that
\begin{equation}
\hat{W}(x) = g x\, W^{-1}(x)\quad\quad \text{when }|x| \leq W(a(g)),
\end{equation}
where $W^{-1}(\cdot)$ is the functional inverse of $z\to W(z)$.
Notice that the position of the cut in this simple disk amplitude is now determined by $W(a)$ which when $a\to 1$ scales as
\begin{align}
g W(a(g)) &= {_2F_1}(1,3/2-s,3/2,a^2)\frac{a}{2} \\
&= \frac{1}{4(s-1)} - \frac{1-a^2}{2(2-s)} + \frac{\sqrt{\pi}\Gamma(1-s)}{4\Gamma(3/2-s)}(1-a^2)^{s-1} + \ldots\label{wsimpa}\\
&= \frac{1}{4(s-1)} - \frac{(1-g^2/g_*^2)^{\frac{1}{s-1/2}}}{2(2-s)} + \frac{\sqrt{\pi}\Gamma(1-s)}{4\Gamma(3/2-s)}(1-g^2/g_*^2)^{\frac{s-1}{s-1/2}} + \ldots
\end{align}
Which of the two last terms dominates depends on whether $s>2$ or $s<2$ (we will not discuss integer $s$). 
In particular, if one identifies the ``simple'' boundary cosmological constant $\hat{\mu}_B$ in analogy with the discussion above \eqref{jz3} one obtains
\begin{equation}
\hat{\mu}_B \sim \begin{cases}
(1-g/g_*)^{\frac{1}{s-1/2}} &\text{for }s>2\\
(1-g/g_*)^{\frac{s-1}{s-1/2}} &\text{for }s<2.\\
\end{cases}
\end{equation}
Defining a corresponding bulk cosmological constant $\hat{\mu}\sim\hat{\mu}_B^2$ and requiring $F(g)\Big|_{\rm singular}\sim \hat{\mu}^{2-\hat{\gamma}_A}$, one gets exactly 
\begin{equation}
\hat{\gamma}_A = \begin{cases}
2-s &\text{for }s>2\\
(s-2)/(s-1) &\text{for }s<2,
\end{cases}
\end{equation}
which is invariant under $s\to s/(s-1)$ and corresponds to the ``right'' branch of \eqref{jz6}.

Let us now have a look at the continuum limit of the simple disk amplitude using \eqref{43}.
Based on \eqref{wsimpa} one expects that one should scale $x^2\to x_c^2(1 - X \epsilon^\beta)$ with $x_c=(s-1/2)/(s-1)$ and $\beta = 1$ for $s>2$ and $\beta = s-1$ for $s<2$, in addition to $a^2\to 1-\sqrt{\Lambda}\epsilon$.
If we denote the leading order of $W^{-1}(x)$ in $\epsilon$ by $W^{-1}(x)\sim 1 + P \epsilon/2$ then for $P>0$
\beq
x = W(1+P\epsilon/2) = x_c - \epsilon P \frac{s-1/2}{2s-4} + \text{analytic} + \epsilon^{s-1}W_\Lambda(P) + \ldots
\eeq
with\footnote{$W_\Lambda(P)$ differs slightly from both $W_{\rm cont}(P)$ and 
$W_{\rm cont}(P)-\tV_{\rm cont}(P)/2$ appearing on the rhs of \rf{43} because 
(1) it is defined
without the factor $g$ multiplying $W(z)$ on the lhs of \rf{43} and (2) it is defined
as the part of $W(z)$ which scales as $\epsilon^{s-1}$. When making the 
substitution $z= 1+\epsilon P$ in $\tV(z)$ in the lhs of \rf{43} we obtain
such a term which together with the appropriately normalized rhs of \rf{43} constitute 
$W_\Lambda(P)$. }
\beq\label{Wcontpos}
W_\Lambda(P):=\frac{\Gamma(1-s)\Gamma(s+1/2)}{\sqrt{\pi}} P^{s-1} -(s-\frac12)\sqrt{\Lambda}^{s-1/2}\frac{\sqrt{P+\sqrt{\Lambda}}}{P}{_2F_1}\Big(1,s;s+\frac12;-\frac{\sqrt{\Lambda}}{P}\Big)
\eeq
It follows that for $s>2$ we have $P = (s-2)/(s-1) X + \epsilon^{s-2}\frac{2s-4}{s-1/2}W_\Lambda\left(\frac{s-2}{s-1}X\right)+\ldots$ and therefore
\begin{align}
\frac{1}{gx}\hat{W}(x) = 1 +\text{analytic} + \epsilon^{s-1} \frac{s-2}{s-1/2} W_\Lambda\left(\frac{s-2}{s-1}X\right)+\ldots,
\end{align}
which has the same form (up to rescaling) as the continuum limit of the non-simple disk function $W(z)$.
On the other hand, when $s<2$ one may check that $W_{\Lambda}(P)$ is monotonically decreasing on $P\in[-\sqrt{\Lambda},\infty\mathclose{[}$ and therefore we identify $P = W_{\Lambda}^{-1}(-x_cX/2) + \mathcal{O}(\epsilon^{2-s})$. This implies that
\beq\label{whatexp}
\frac{1}{gx} \hat{W}(x) = 1 + \epsilon W_{\Lambda}^{-1}(-x_cX/2) + \mathcal{O}(\epsilon^{3-s}).
\eeq
Since we took $x^2\to x_c^2(1-X\epsilon^{s-1})$, the linear term in (\ref{whatexp}) is in fact the dominant singular part and we conclude that it is really the functional inverse of \eqref{Wcontpos} that provides the continuum limit of the simple disk amplitude.

\section{Relation to the multicritical $O(n)$ loop models}\label{Onmodel}

The range of universality classes parametrized by $s\in\mathopen{]}1,\infty\mathclose{[}$ is akin to that of the multi-critical $O(n)$ models studied in \cite{KS}.
This is not a coincidence: it has been observed in \cite{MLeG11,BBG12} that at criticality there exists a natural relation between $O(n)$ models and random planar maps with non-trivial weights on the faces. 
Let us briefly describe this connection.

\begin{figure}
\centering
\includegraphics[width=.9\linewidth]{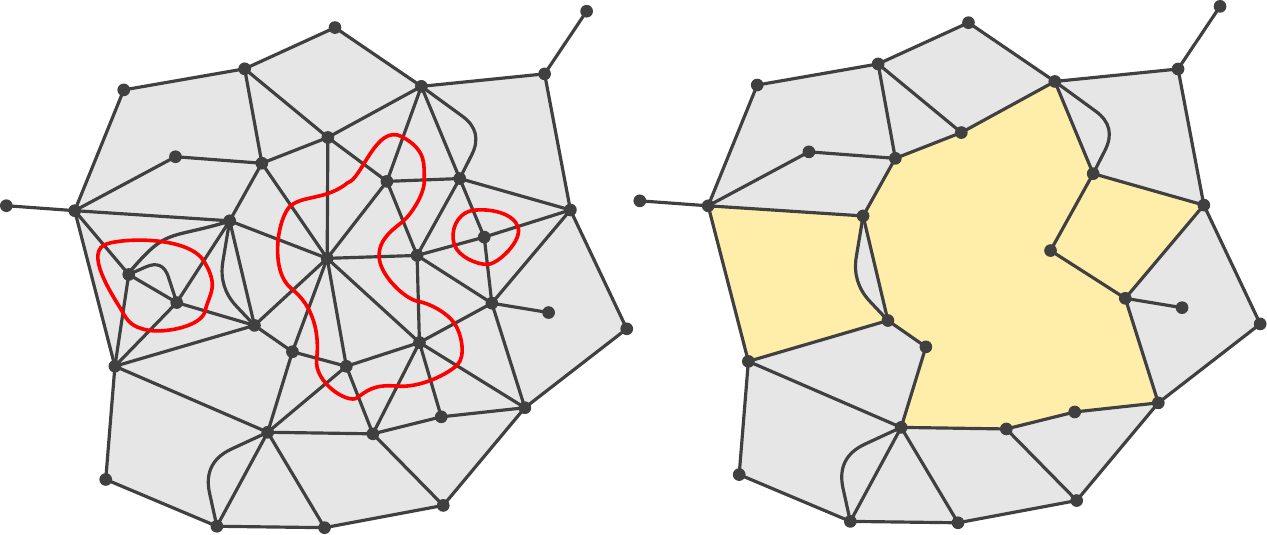}
\caption{An example of a loop-decorated planar map with a boundary (left) and its gasket (right).}
\label{fi:gasket}
\end{figure} 

The $O(n)$ matrix model for positive integer $n$ is defined by
\beq
Z = \int d M \prod_{i=1}^n d\Phi_i \exp\left( -N \tr\left[ \bar{V}(M) + \frac12 \sum_{i=1}^n \Phi_i^2 - \frac{1}{2 z_*}\sum_{i=1}^n \Phi_i^2 M\right]\right),
\eeq
where $\bar{V}(M) = \frac12 M^2 - \sum_{k=2}^{m+1}\frac{1}{k}\bar{q}_k M^k$ is some polynomial potential and $z_*$ is an independent coupling constant.
The corresponding Feynman diagrams in the large $N$ limit can be interpreted as the duals of loop-decorated planar maps, as shown in figure \ref{fi:gasket} (left). 
Each such loop-decorated planar map comes with a weight 
\begin{equation}\label{loopweight}
n^{\#\text{loops}} \left(2z_*\right)^{-\#\text{loop-decorated triangles}} \prod_{\text{non-loop faces }f} \bar{q}_{\deg(f)},
\end{equation}
which makes sense for any real value of $n$.
It is shown in \cite{KS} that for $n \in [-2,2]$ one may tune the parameters $\bar{q}_{k}$, $k = 2,\ldots,m+1$, together with $z_*$ such that the disk amplitude $\bar{W}(z)$ takes the form
\begin{equation}
\bar{W}(z) = C_1\, (z-z_*)^m \cos( b\, \arccos(C_2/(z - z_*))) + \text{polynomial}
\end{equation}
for $z>z_*$, where $b = \arccos(n/2)/\pi \in \mathopen{[}0,1\mathopen{]}$ and $C_1$ and $C_2>0$ are constants.
In particular $\bar{W}(z)|_{\text{sing.}} \sim (z-z^*)^{m-b}$ when $z\to z^*$.
One should notice that this is precisely the scaling of the generalized Kazakov disk amplitude $W(z)$ at the critical value $g=g_*$ when $s=m-b+1$, since at $g=g_*$ we have explicitly
\begin{equation}
W(z) = \frac{1}{z}{_2F_1}(1/2,1;1/2+s;z^{-2}),\quad W(z)|_{\text{sing.}} \sim (z-1)^{s-1}.
\end{equation} 
Of course, when $n=0$, i.e. $b=1/2$, the loops are suppressed and one is back at the standard multi-critical matrix model (although with a different potential than the one in Sec. \ref{multicritical} since it is not restricted to be even).

To better understand the connection between the two models it is convenient (see also \cite{BBG12}) to introduce the \emph{gasket} $\mathcal{G}(m)$ of a loop-decorated planar map $m$ with a boundary to be the planar map obtained by removing all triangles intersected or surrounded by loops (see figure \ref{fi:gasket}). 
Given a planar map $m'$ with boundary (and no loops), one may ask for the total weight in the sense of (\ref{loopweight}) of all loop-decorated planar maps $m$ that have $m'=\mathcal{G}(m)$ as their gasket.
Since each face of $m'$ corresponds to either a face or a loop of $m$, we easily find that this total weight factorizes as $\prod_{f\in\text{Faces}(m')} q_{\deg(f)}$ where $q_k$ is given by
\begin{equation}
q_k := \bar{q}_{k} + n \sum_{l=0}^\infty \binom{l+k}{l} (2z_*)^{-l-k} \bar{W}^{(l)}.
\end{equation}
This is precisely the weight associated to $m'$ in a one-matrix model with ``effective'' potential
\begin{align}
V_{\text{eff}}'(z) &:= z - \sum_{k=1}^\infty q_k z^{k-1}\\
& = \bar{V}'(z) - n \sum_{k=1}^\infty\sum_{l=0}^{\infty} z^{k-1} \binom{l+k}{l} (2z_*)^{-l-k} \bar{W}^{(l)} \\
&= \bar{V}'(z) - n \frac{2z_*}{z} (\bar{W}(2z_*-z)-\bar{W}(2z_*)) 
\end{align}
and therefore $\bar{W}(z)$ may be identified as the disk amplitude of this one-matrix model.
In particular the singular behavior $V_{\text{eff}}'(z)|_{\text{sing.}} \sim (z_* - z)^{m-b}$ agrees with that of (\ref{Vmult}) when $s=m-b+1$.

The precise connection with the multi-critical $O(n)$ model only holds at criticality, i.e. $g=g_*$. 
This explains why the continuum limit (\ref{ja54}) of our disk amplitude for $\Lambda \neq 0$ is quite different from the standard one of the $O(n)$ model which reads \cite{KS}
\begin{equation}
W_{O(n)}(P) \propto \cosh( (m-b)\,\text{arccosh}(P/\sqrt{\Lambda})). 
\end{equation}
Note that one can obtain a connection away from criticality, i.e. $g<g_*$, if one is willing to supplement the $O(n)$ model weight (\ref{loopweight}) with a factor $g/g_*$ for each vertex in the gasket, i.e. for each vertex not surrounded by a loop.

\section{Conclusions}\label{discuss}

We have shown that standard matrix model 
calculations extend to potentials of the form 
\beq\label{ja60}
V(x) = \frac 1g \sum_n v_n x^{2n}, \quad v_n \sim \frac{1}{n^{s+1} }
\quad {\rm for} \;\;  n \to \infty.
\eeq
Both the potential and their derivatives have cuts on the real axis.
Nevertheless one can find 1-cut solutions $W(z)$ to the disk amplitude
which are  natural generalizations of the standard multicritical disk 
amplitudes and in this way the generalized Kazakov potentials 
$V_s(x)$ serve as generalized multicritical points interpolating  between the 
standard multicritical points.
In particular the b-loop functions are universal functions when expressed 
in terms of   $z_j^2-a^2$, $j=1,\ldots,b$ and the $b-2$ first moments 
$M_k$, $k=1,\ldots,b-2$, even if $W(z)$ itself depends on infinite many $M_k$'s.
Also, for the multiloop functions the continuum limit is obtained in a straigth forward
manner. 

To each $s > 1$ one can formally associate a central change $c(s)$ given
by \rf{jz6} and conversely
to each central charge $c <1$ one can associate two values $s(c) >2$ and $s'(c) < 2$
related by $s' = s/(s-1)$ corresponding the  KPZ exponents 
$\gamma_A (s) = 2-s$ and $\gamma'_A(s')=2-s'$, related by \rf{jz7} and corresponding
to the two solutions of the KPZ equation \rf{jz6}. The ``wrong''  solution of the
KPZ equation where $\gamma_A'(s') > 0$ has been associated with so-called
touching interactions where one in matrix model context has added terms
like $g_t(\tr \phi^2)^2$ to the ordinary matrix potential. By fine-tuning 
the touching coupling like $g_t$ one could obtain certain critical exponents $\gamma > 0$.
We have here seen very explicitly in Sec.\ \ref{combinatorial} that for potentials 
with the most heavy tail, namely $1 < s < 2$ the ``touching'' picture appears
automatically, without adding any explicit touching interaction, and that the 
whole range $0 < \gamma_A' < 1$ is spanned.

A number of interesting questions remain to be answered. Is there any conformal 
field theory interpretation of the region $1/2 < s < 1$? How do the perturbations
away from the generalized Kazakov point relate to  the corresponding 
conformal field theory, which in general will be irrational? What {\it is} the most natural
way to perturb away from the generalized Kazakov point and how does it relate
to the standard KdV flow equations valid for any standard multicritical model?
These questions deserve further considerations.

\subsection*{Acknowledgments}

The authors warmly thanks Leonid Chekhov for numerous discussions
and they acknowledge  the support by  the ERC-Advance
grant 291092, ``Exploring the Quantum Universe'' (EQU).
Y.~M.\ thanks the Theoretical Particle Physics and Cosmology group 
at the Niels Bohr Institute for the hospitality.

\section*{Appendix A: Simplest example of logarithmic potential \label{s:slog}}

We illustrate in this Appendix how general formulas of Sect.~\ref{riemann-hilbert} work
for potentials which are  not even, \ie $V(x)\neq V(-x)$.
A simplest such a potential for which $V'$ has a cut at the real axis is the logarithmic 
potential
\be
V(x)=\frac 1g \left[(1-x){\log}(1-x)+x \right]=
\frac 1g  \sum_{n=1}^\infty \frac{x^{n+1}}{n(n+1)} ,
\label{VV3}
\ee
so that both $V(x)$ and
\be
V'(x)=-\frac1g\log (1-x) =\frac 1g  \sum_{n=1}^\infty \frac{x^n}{n} 
\label{V3}
\ee
have a cut from $1$ to $\infty$.

From \eq{Wgen} we find the solution
\be
W(z)=\frac1{g}
\left[\arctanh \sqrt{\frac{(z-b)}{(z-a)}} -\arctanh \sqrt{\frac{(1-a)(z-b)}{(1-b)(z-a)}}
-\frac 12 \log(1-z)
\right],
\label{W3}
\ee
where
\be
a=b-4\left(  1-\sqrt{1-b} \right)
\label{b-a}
\ee
and
\be
g=\frac{(b-a)^2}{16}= \left(  1-\sqrt{1-b} \right)^2.
\label{g3}
\ee
The cut $[a,b]$ is non-symmetric. 

The solution \rf{W3} has all required properties: it is analytic outside of $[a,b]$,
 reproduces Wigner's law
as $g\to 0$ etc.. The discontinuity across the cut determines the 
(normalized) spectral density
\be
\rho(x)=\frac1{\pi g}
\left[\arctan \sqrt{\frac{(1-a)(b-x)}{(1-b)(x-a)}}-\arctan \sqrt{\frac{(b-x)}{(x-a)}} 
\right],
\label{rho3}
\ee
which indeed
obeys \eq{sp} with the potential \rf{V3} as can be explicitly checked. 
The spectral density  \rf{rho3} is positive for $b<1$, 
vanishes at the ends of the cut, but looks pretty
different from the previously known cases, where $V'$ has no cut at the real axis.
In those usual cases $\rho$ has a square-root singularity, which is now hidden
under the arctan. 

A critical behavior is now reached as $b\to1$, when
\be
g \to g_*-2 \sqrt{1-b}, \qquad g_*=1
\label{24}
\ee 
from \eq{g3}. Expanding near the critical point similarly to \rf{ja50},
\be
b=1-\epsilon \sqrt{\Lambda},\qquad z=1+\epsilon P,
\ee
we find from \eq{W3} 
\be
W-\frac {V'}2 \propto \arctanh \sqrt{\frac P{\sqrt{\Lambda}}+1}
\label{S1}
\ee
which has a cut along the real axis for $p<-\sqrt{\Lambda}$
and
\be
\rho_{\rm cont.} (p)=
\frac1{\pi} \arctanh {\sqrt{-1-p/\sqrt{\Lambda}}}.
\ee
Notice that $\epsilon$ has canceled on the right-hand side of \eq{S1}.
This might imply that the double scaling limit does not exist for this matrix model,
but it rather corresponds to a certain continuum 
combinatorial problem like the Kontsevich matrix model.
A similar behavior occurs for the potential \rf{VV5} for $s=1$.
The potentials \rf{VV3} and \rf{VV5} with $s=1$ thus belong 
to the same universality class.

\section*{Appendix B: An extension of the polylogarithm\label{appa}}

The polylogarithm has the integral representation
\be
{\rm Li}_s (A)=\sum_{n=1}^\infty \frac{A^n}{n^s}=
\frac1{\Gamma(s)} \int _0^\infty \d \tau\, \tau^{s-1}
\left( \frac 1{1-A \e^{-\tau}}-1\right),
\label{integ}
\ee
where the integral is convergent at small $t$ for $s>0$ and $|A|<1$
and $s>1$ for $A=1$..
The asymptotic behavior of ${\rm Li}_s (A)$ as $A\to 1$ depends on
the value of $s$. For $0<s<2$ we have from \eq{integ} 
\be
{\rm Li}_s (A) \to\zeta(s)+\Gamma(1-s) \left( 1-A\right)^{s-1} 
\qquad \hbox{for}~~0<s<2
\label{aa1}
\ee
and
\be
{\rm Li}_s (A) \to\zeta(s)+\zeta(s-1) \left( 1-A\right)
\qquad \hbox{for}~~s>2.
\label{aa2}
\ee

Let us define the function
\be
F_s\left(\alpha,A\right)
= \sum_{n=1}^\infty \frac{\Gamma\left(\alpha+n\right)}
{\Gamma\left(\alpha+1\right)n!} \frac{A^n}{n^s}
 = \frac1{\alpha\Gamma(s)}
\int_0^\infty \d \tau \,\tau^{s-1}\left[ \frac1{\left({1-A \e^{-\tau}}\right)^\alpha}-1 \right]
\label{defFa}
\ee
which for $\alpha=1$ reduces to the polylogarithm and for $\alpha=1/2$ 
reproduces the function \rf{defF}.
The integral in \eq{defFa} is convergent for $s>0$ if $|A|<1$ and 
$s>\alpha$ if $A=1$. The derivative of \rf{defFa} reads
\be
\frac{\d}{\d A}F_s\left(\alpha,A\right) = \frac 1A F_{s-1}\left(\alpha,A\right) .
\label{A5}
\ee

The asymptotic behavior near $A=1$ can be found from the difference
\be
F_s\left(\alpha,A\right) - F_s\left(\alpha,1\right)= \frac1{\alpha\Gamma(s)}
\int_0^\infty \d \tau \,\tau^{s-1}\left[ \frac1{\left({1-A \e^{-\tau}}\right)^\alpha}-
\frac1{\left({1-\e^{-\tau}}\right)^\alpha} \right].
\label{difFa}
\ee
If we expand the difference in $(1-A)$, we find
\be
F_s\left(\alpha,A\right) - F_s\left(\alpha,1\right)= -\frac{(1-A)}{\Gamma(s)}
\int_0^\infty \d \tau \,\tau^{s-1}\left[ \frac{\e^{-\tau}}{\left({1-\e^{-\tau}}\right)^{\alpha+1}} \right],
\label{difFa1}
\ee
where the integral converges for $s>1+\alpha$. We then find
\be
F_s \left(\alpha,A\right) - F_s \left(\alpha,1\right) =
- \left( 1-A\right)F_{s-1}\left(\alpha,1\right)
\qquad \hbox{for}~~s>1+\alpha.
\label{aaa2}
\ee
If $s<1+\alpha$, the integral in \eq{difFa1} diverges as $\tau\to 0$
and we cannot expand in $(1-A)$. Then for $\alpha<s<1+\alpha$ 
the right-hand side of \eq{difFa} is
dominated by small $\tau \sim (1-A)$ and we write
\bea
F_s\left(\alpha,A\right) - F_s\left(\alpha,1\right)&= &\frac1{\alpha\Gamma(s)}
\int_0^{\sim 1} \d \tau \,\tau^{s-1}\left[ \frac1{\left({1-A +\tau}\right)^\alpha}-
\frac1{\tau^\alpha} \right] 
\non &=& \left(  1-A  \right)^{s-\alpha}
 \frac{\Gamma(\alpha-s)}{\Gamma(1+\alpha)}
\qquad  \hbox{for} ~~\alpha<s<1+\alpha .
\label{difFa2}
\eea
For $\alpha<s<1+\alpha$ 
this is larger than the contribution from the domain of large $\tau$, where
\be
\int_{\sim 1}^\infty \d \tau \,\tau^{s-1}
\left[ \frac1{\left({1-A \e^{-\tau}}\right)^\alpha}-
\frac1{\left({1-\e^{-\tau}}\right)^\alpha} \right]
 \propto - \left(  1-A  \right) .
\ee
Thus the asymptote \rf{difFa2} holds for  $\alpha<s<1+\alpha$ and
the asymptote \rf{aaa2} holds for  $s>1+\alpha$.


\begin{thebibliography}{99}

\bibitem{adf}
J.~Ambjorn, B.~Durhuus and J.~Frohlich,
 {\it Diseases of triangulated random surface models, and possible cures,}
  Nucl.\ Phys.\ B {\bf 257} (1985)  433;
  {\it The appearance of critical dimensions in regulated string theories. 2},
  Nucl.\ Phys.\ B {\bf 275}, 161 (1986).
J.~Ambjorn, B.~Durhuus, J.~Frohlich and P.~Orland,
  {\it The appearance of critical dimensions in regulated string theories},
  Nucl.\ Phys.\ B {\bf 270}, 457 (1986).

  
\bibitem{david1}
F.~David,
{\it A model of random surfaces with nontrivial critical behavior},
  Nucl.\ Phys.\ B {\bf 257}, 543 (1985).
  A.~Billoire and F.~David,
  {\it Scaling properties of randomly triangulated planar random surfaces: A numerical study},
   Nucl.\ Phys.\ B {\bf 275}, 617 (1986).

\bibitem{kkm}
 V.~A.~Kazakov, A.~A.~Migdal and I.~K.~Kostov,
{\it Critical properties of randomly triangulated planar random surfaces,}
  Phys.\ Lett.\ B {\bf 157}, 295 (1985).
  \bibitem{kkm1}
  D.~V.~Boulatov, V.~A.~Kazakov, I.~K.~Kostov and A.~A.~Migdal,
  {\it Analytical and numerical study of the model of dynamically 
  triangulated random surfaces,}
  Nucl.\ Phys.\ B {\bf 275}, 641 (1986).


\bibitem{david2}
 F.~David,
{\it Planar diagrams, two-dimensional lattice gravity and surface models,}
  Nucl.\ Phys.\ B {\bf 257}, 45 (1985);
  


\bibitem{kazakov2}
  V.~A.~Kazakov,
  {\it The appearance of matter fields from quantum fluctuations of 2D gravity,}
  Mod.\ Phys.\ Lett.\ A {\bf 4} (1989) 2125.



\bibitem{staudacher}
  M.~Staudacher,
{\it The Yang-Lee edge singularity on a dynamical planar random surface,}
  Nucl.\ Phys.\ B {\bf 336} (1990) 349.


\bibitem{review}
  P.~Di Francesco, P.~H.~Ginsparg and J.~Zinn-Justin,
  {\it 2-D Gravity and random matrices,'}
  Phys.\ Rept.\  {\bf 254} (1995) 1


\bibitem{kazkov}
  J.~M.~Daul, V.~A.~Kazakov and I.~K.~Kostov,
  {\it Rational theories of 2-D gravity from the two matrix model,}
  Nucl.\ Phys.\ B {\bf 409} (1993) 311,
  [hep-th/9303093].

\bibitem{Kos89}
 I.~K.~Kostov, {\it O(N) vector model on a planar random lattice: spectrum of
anomalous dimensions,}
Mod.\ Phys.\ Lett.\ A {\bf 4} (1989) 217.

\bibitem{KS}
 I.~K.~Kostov and M.~Staudacher,
{\it Multicritical phases of the O(n) model on a random lattice,}
Nucl.\ Phys.\ B {\bf 384} (1992) 459  [hep-th/9203030].


\bibitem{MLeG11}
  J.-F.~Le~Gall and G.~Miermont,
  {\it Scaling limits of random planar maps with large faces,}
  Ann.\ Probab. {\bf 39} (2011) 1--69
  [arXiv:0907.3262]

\bibitem{BBG12}
  G.~Borot, J.~Bouttier and E.~Guitter,
  {\it A recursive approach to the O(n) model on random maps via nested loops,}
  J.\ Phys.\ A {\bf 45} (2012) 045002
  [arXiv:1106.0153].

\bibitem{BC16}
  T.~Budd, N.~Curien,
  {\it Geometry of infinite planar maps with high degrees,}
   [arXiv:1106.0153].  

\bibitem{BIPZ}
  E.~Br\'ezin, C.~Itzykson, G.~Parisi and J.B.~Zuber,
  {\it Planar Diagrams,}
  Commun.\ math.\ Phys.\ 59 (1978) 35  

\bibitem{GG95}
  R.~Gopakumar and G.J.~Gross,
  {\it Mastering the master field,}
  Nucl.\ Phys.\ {\bf B451} (1995) 379
  [hep-th/9411021]

\bibitem{durhuus}
  B.~Durhuus,
  {\it Multispin systems on a randomly triangulated surface,}
  Nucl.\ Phys.\ B {\bf 426} (1994) 203,
  [hep-th/9402052].

\bibitem{adj}
  J.~Ambjorn, B.~Durhuus and T.~Jonsson,
  {\it A Solvable 2-d gravity model with gamma $> 0$,}
  Mod.\ Phys.\ Lett.\ A {\bf 9} (1994) 1221,
    [hep-th/9401137].

\bibitem{klebanov}
  I.~R.~Klebanov,
  {\it Touching random surfaces and Liouville gravity,}
  Phys.\ Rev.\ D {\bf 51} (1995) 1836,
  [hep-th/9407167].\\
  J.~L.~F.~Barbon, K.~Demeterfi, I.~R.~Klebanov and C.~Schmidhuber,
 {\it Correlation functions in matrix models modified by wormhole terms,}
  Nucl.\ Phys.\ B {\bf 440} (1995) 189,
  [hep-th/9501058].


\bibitem{book}
 J.~Ambjorn, B.~Durhuus and T.~Jonsson,
 {\it Quantum geometry. A statistical field theory approach,}
  Cambridge  (UK) Univ.\ Press (1997).


\bibitem{am}
 J.~Ambjorn and Y.~M.~Makeenko,
{\it Properties of loop equations for the Hermitean matrix model and for two-dimensional
 quantum gravity,}
  Mod.\ Phys.\ Lett.\ A {\bf 5} (1990) 1753.

\bibitem{ajm}
 J.~Ambjorn, J.~Jurkiewicz and Y.~M.~Makeenko,
{\it Multiloop correlators for two-dimensional quantum gravity,}
  Phys.\ Lett.\ B {\bf 251} (1990) 517.

\bibitem{ackm}
  J.~Ambjorn, L.~Chekhov, C.~F.~Kristjansen and Y.~Makeenko,
  {\it Matrix model calculations beyond the spherical limit,}
  Nucl.\ Phys.\ B {\bf 404} (1993) 127
   Erratum: [Nucl.\ Phys.\ B {\bf 449} (1995) 681],
  [hep-th/9302014].
  
  \bibitem{to-appear}
 J.~Ambjorn, T.~Budd, L.~Chekhov  and Y.~Makeenko,
 {\it to appear} 

\end{thebibliography}
\end{document}